\documentclass[twocolumn]{openjournal}
\usepackage{natbib}
\usepackage{graphicx,amsmath,amssymb,amstext}
\usepackage{amsbsy,amsfonts,amsthm,color}
\usepackage[colorlinks,linkcolor=blue,citecolor=blue,urlcolor=blue ]{hyperref}
\usepackage[utf8]{inputenc}
\usepackage{float}
\usepackage[caption=false]{subfig}
\usepackage{upgreek}

\newcommand{\eagle}{\mbox{EAGLE}}

\newcommand{\flares}{\mbox{FLARES}}

\newcommand{\Mbh}{M_{\bullet}}
\newcommand{\Msun}{{\rm M_{\odot}}}

\defcitealias{lovell2020}{\flares\,I}

\newcommand\blfootnote[1]{%
  \begingroup
  \renewcommand\thefootnote{}\footnote{#1}%
  \addtocounter{footnote}{-1}%
  \endgroup
}

\begin{document}

\title{First Light and Reionization Epoch Simulations (FLARES) - XV: The physical properties of super-massive black holes and their impact on galaxies in the early Universe\vspace{-3em}}

\author{Stephen M. Wilkins$^{1\star\dagger}$} 
\author{Jussi K. Kuusisto$^{1\dagger}$} 

\author{Dimitrios Irodotou$^{2}$} 
\author{Shihong Liao$^{3}$} 
\author{Christopher C. Lovell$^{4}$} 
\author{Sonja Soininen$^{2}$} 

\author{Sabrina Berger$^{5,6}$}
\author{Sophie L. Newman$^{4}$}
\author{William J. Roper$^{1}$} 
\author{Louise T. C. Seeyave$^{1}$} 
\author{Peter A. Thomas$^{1}$} 
\author{Aswin P. Vijayan$^{1,7,8}$} 
\blfootnote{$^{\dagger}$ Joint primary authors.}
\blfootnote{$^{\star}$ Corresponding author. Email: \href{mailto:s.wilkins@sussex.ac.uk}{s.wilkins@sussex.ac.uk}}

\affiliation{$^1$Astronomy Centre, University of Sussex, Falmer, Brighton BN1 9QH, UK}
\affiliation{$^2$The Institute of Cancer Research, 123 Old Brompton Road, London SW7 3RP, UK}
\affiliation{$^3$Key Laboratory for Computational Astrophysics, National Astronomical Observatories, Chinese Academy of Sciences, Beijing 100101, China}
\affiliation{$^4$Institute of Cosmology and Gravitation, University of Portsmouth, Burnaby Road, Portsmouth, PO1 3FX, UK}

\affiliation{$^5$School of Physics, University of Melbourne, Parkville, VIC 3010, Australia}
\affiliation{$^6$ARC Centre of Excellence for All Sky Astrophysics in 3 Dimensions (ASTRO 3D). Australia}
\affiliation{$^7$Cosmic Dawn Center (DAWN)}
\affiliation{$^8$DTU-Space, Technical University of Denmark, Elektrovej 327, DK-2800 Kgs. Lyngby, Denmark}

\begin{abstract}

Understanding the co-evolution of super-massive black holes (SMBHs) and their host galaxies remains a key challenge of extragalactic astrophysics, particularly the earliest stages at high-redshift. However, studying SMBHs at high-redshift with cosmological simulations, is challenging due to the large volumes and high-resolution required.  Through its innovative simulation strategy, the First Light And Reionisation Epoch Simulations (FLARES) suite of cosmological hydrodynamical zoom simulations allows us to simulate a much wider range of environments which contain SMBHs with masses extending to $\Mbh>10^{9}\ \Msun$ at $z=5$. In this paper, we use FLARES to study the physical properties of SMBHs and their hosts in the early Universe ($5\le\, z \le10$). FLARES predicts a sharply declining density with increasing redshift, decreasing by a factor of 100 over the range $z=5\to 10$.  Comparison between our predicted mass and bolometric luminosity functions and pre-\emph{JWST} observations yielded a reasonable match. However, recent \emph{JWST} observations appear to suggest a higher density of SMBHs at $z\approx 5$ and the presence of more luminous SMBHs at $z>6$ than predicted by FLARES. Finally, by using a re-simulation with AGN feedback disabled, we explore the impact of AGN feedback on their host galaxies. This reveals that AGN feedback results in a reduction of star formation activity, even at $z>5$, but only in the most massive galaxies. A deeper analysis reveals that AGN are also the cause of suppressed star formation in passive galaxies but that the presence of an AGN doesn't necessarily result in the suppression of star formation.

\end{abstract}


\maketitle



\section{Introduction}\label{sec:intro}
Since the first conceptualization of black holes almost 250 years ago by John Mitchell and Pierre--Simon Laplace \citep[see e.g.][]{Schaffer79,Montgomery2009} and the first observation of a quasar 60 years ago by \cite{Schmidt1963}, numerous scientists have worked on understanding how black holes form, evolve, and affect their surroundings \citep[some of the most seminal works include][]{Kerr1963,Salpeter1964,Penrose1965,LyndenBell1969,Penrose1971,Bardeen1973,Shakura1973,Blandford1977,Abramowicz1988,Narayan1994}. However, a complete theory of how black holes operate and interact with their host galaxies \citep[e.g.][]{Rees1984,Richstone1998} remains one of the biggest challenges in modern (astro)physics today.

Supermassive black holes (SMBHs) with masses ranging from $\sim$$10^{6}\ \Msun$ to $\sim$$10^{10}\ \Msun$ have been observed to lie in the centres of massive galaxies \citep{Kormendy1995} and to follow tight correlations with their host galaxy properties \citep[e.g.][]{Magorrian1998,Ferrarese2000,Gebhardt2000,Tremaine2002,Marconi2003,Merloni2003,Haring2004,McConnell2013}. Therefore, understanding the co--evolution of SMBHs and their hosts is an essential part of galaxy formation theory \citep{Silk1998,Kauffmann2000,Kormendy2013}.

Observations of high redshift quasars \citep[e.g.][]{Jiang2016, Matsuoka2016, Maiolino2023a} have revealed that SMBHs existed in the Universe less than a billion years after the Big Bang \cite[see][for recent reviews]{Inayoshi2020,Fan2023}.
The traditional Eddington-limited stellar remnant BH formation scheme cannot explain such massive BHs in the early Universe; in order to grow BHs up to $\sim10^{9}\ \Msun$ at $z \gtrsim 5$, alternative formation mechanisms are required, such as massive seeds and/or enhanced BH accretion \citep{Latif2013,Volonteri2021}. Suggested massive seed formation scenarios include direct collapse black holes and the collapse of very massive stars formed through mergers \citep{Loeb1994,Madau2001,Bromm2003,PortegiesZwart2004,Volonteri2005,Begelman2006,Regan2009}. Understanding not only the formation but also the evolution and physics of high redshift SMBHs is essential in order to capture the effects that SMBHs have on their surroundings. Since SMBHs grow by accreting surrounding gas while simultaneously releasing energy into it \citep[e.g.][]{Fabian2012}, they affect their surroundings, thus altering the overall properties of their host galaxies \citep[e.g. the bright end of the galaxy luminosity function][]{Kauffmann2000,Granato2004,Bower2006,Cattaneo2006,Croton2006}.

In addition to influencing its host galaxy, the radiation emitted in the vicinity of black holes also contributes to the overall ionizing photon budget, although stellar sources of ultraviolet photons seem to predominantly drive the hydrogen reionization of the Universe \citep{Madau2015,Qin2017,Dayal2018,Robertson2022}.
Since the number density of Active Galactic Nuclei (AGN) increases rapidly towards lower redshifts, the fractional contribution of AGN to the total ionizing photon budget becomes more significant towards lower redshifts. Although their ionizing photon contribution during hydrogen reionization was not dominant, their higher number density and harder spectra suggest that AGN significantly contributed to helium reionisation and to the meta-galactic UV and X-ray background of the Universe \citep[e.g.][]{Ricotti2004,Giallongo2019,Puchwein2019,Finkelstein2022}.

From a theoretical/computational point of view, black hole physics has been an integral component of models of galaxy formation, which try to capture the effects of black hole feedback on the simulated galaxies \citep[see the reviews of][]{Somerville2015,Naab2017,Vogelsberger2020, Habouzit2022a, Habouzit2022b}. Since SMBHs grow by accreting surrounding gas while simultaneously releasing energy to it \citep[e.g.][]{Fabian2012}, they affect their surroundings thus altering the overall properties of their host galaxies \citep{Ciotti2001, DiMatteo2005, Murray2005, Hopkins2006}. Traditionally, black hole feedback has been incorporated either through a thermal or quasar mode, where a fraction of the bolometric luminosity is injected as thermal energy to the surrounding environment \citep{Springel2005,Booth2009,Tremmel2017} or as kinetic mode \citep[][]{Croton2006,Costa2014,Choi2015,Costa2020}, or as a combination of different modes \citep{Sijacki2007,Dubois2012,Sijacki2015,Weinberger2017,Dave2019}. However, different implementations of black hole physics result in discrepancies in the predictions of black hole properties both at low and at high redshifts \citep{Meece2017,Habouzit2022a,Habouzit2022b}, which makes understanding the co--evolution of black holes and galaxies even more challenging.

With the advent of \emph{JWST} the observational SMBH frontier is now shifting to higher-redshift. Samples of SMBHs have now been detected out to $z\approx 10$, deep into the Epoch of Reionisation \citep{Larson2023, Harikane2023, Juodzbalis2023, Matthee23, Greene23, Kocevski23, Maiolino2023a, Ubler2023, Kokorev2024} with tentative detections at $z>10$ \citep{Maiolino2023b, Bogdan2023, Juodzbalis2023}. The innovation of \emph{JWST} is its ability to constrain AGN activity in galaxies through broad line emission, line-ratios, compact morphology, broad-band photometry, or a combination thereof. With new imaging and spectroscopic surveys underway, or planned, samples of high-redshift AGN will inevitably grow in size and robustness. \emph{JWST} observations will also soon be complemented by wide-area observations from \emph{Euclid}, providing large samples of bright, AGN-dominated sources.

The contribution of \emph{JWST}, and soon \emph{Euclid}, represents an important new frontier in cosmological galaxy formation. Comparison between these observations and galaxy formation models will provide the opportunity to constrain the formation and growth mechanisms of SMBHs in the early Universe.

However, simulating large samples of SMBH dominated galaxies in the early Universe is challenging due to their relative rarity, thus requiring large simulations. Flagship simulations such as Illustris \citep{Vogelsberger2014a, Vogelsberger2014b, Genel2014, Sijacki2015}, EAGLE \citep{schaye_eagle_2015, crain_eagle_2015, McAlpine2016, McAlpine2017}, Horizon-AGN \citep{Dubois2016, Volonteri2016}, TNG100 \citep{Weinberger2017, Marinacci2018, Naiman2018, Nelson2018, Pillepich2018b, Pillepich2018a, Springel2018,  Weinberger2018}, Simba \citep{Dave2019}, etc., are too small to yield statistically useful samples of observationally accessible massive SMBHs in the early Universe \citep[see][]{Habouzit2022b}. While larger simulations exist, including BAHAMAS \citep{McCarthy2017}, TNG300 \citep{Weinberger2017, Marinacci2018, Naiman2018, Nelson2018, Pillepich2018b, Pillepich2018a, Springel2018,  Weinberger2018}, FLAMINGO \citep{Schaye2023}, most have significantly lower mass-resolution, limiting their use for studying the SMBHs now accessible to \emph{JWST}. The exceptions are simulations that only target the high-redshift Universe, for example: Massive Black \citep{Khandai2012}, Bluetides \citep{DiMatteo2017, Wilkins2017, Tenneti2018, Huang2018, Ni2020, Marshall2020}, ASTRID \citep{Bird2022, Ni2022}, and more recently the First Light And Re-ionisation Epoch Simulations \citep[FLARES,][]{FLARES-I,FLARES-II}, the focus of this study.

In this work, we utilise the FLARES suite to study SMBHs in the distant, high-redshift ($5 \le\, z \le\, 10$) Universe. FLARES is a suite of hydrodynamical zoom-in simulations, where a range of different overdensity regions were selected from a large dark matter only periodic volume and re-simulated using a variant of the EAGLE \citep{schaye_eagle_2015, crain_eagle_2015} physics model. The benefit of this simulation strategy is that it allows rare, high-density regions to be simulated with full hydrodynamics and relatively high resolution, without the need to simulate large periodic volumes with full hydrodynamics. The regions can also be statistically combined to produce composite distribution functions, mimicking a larger box. This method allows us to probe statistical distributions of galaxies within a higher effective volume, simulate extremely massive galaxies hosting extremely massive black holes (potential AGN), and test the EAGLE model at high redshift.

This paper is structured as follows: in Section \ref{sec:sim} we detail the simulation suite FLARES as well as modelling methodology for SMBH and stellar emission. In Section \ref{sec:blackholes} we present predictions for the physical and observational properties of SMBHs, including the environmental dependence (\S\ref{sec:blackholes:mf:environment}). In Section \ref{sec:hosts} we explore the correlation of SMBH properties with the properties of their hosts. In Section \ref{sec:impact} we briefly discuss the impact of SMBHs on their host galaxies, and finally in Section \ref{sec:conclusions} we present our conclusions. 

\section{Simulations and Modelling}\label{sec:sim}

In this work we use the First Light And Reionisation Simulations \citep[FLARES,][]{FLARES-I, FLARES-II} to explore predictions for the properties of super-massive black holes (SMBHs) in the early ($5 \le\, z \le\, 10$) Universe. In this section we describe the wider FLARES project and the underpinning EAGLE physical model, focusing on the SMBH physics.

\subsection{FLARES}

\begin{figure}
    \centering
    \includegraphics[width=0.9\columnwidth]{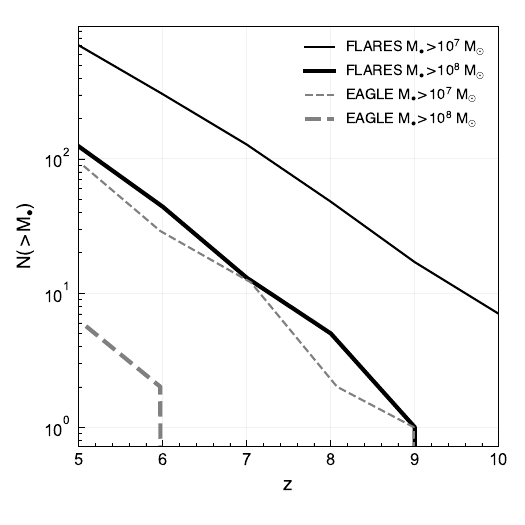}
    \caption{The number of $M_\bullet>10^{7}\ {\rm M_{\odot}}$ (thin line) and $M_\bullet>10^{8}\ {\rm M_{\odot}}$ (thick line) super--massive black holes in \flares\ (solid--line) and the \eagle\ (dashed line) reference volume at $z=5-10$. The FLARES simulation strategy results in the simulating of many more massive SMBHs than the original EAGLE reference simulation despite a comparable total volume.} 
    \label{fig:sim:nbh}
\end{figure}

The core \flares\ simulations are a suite of 40 independent hydrodynamical re-simulations of spherical regions of size $\rm 14\,cMpc\,$$h^{-1}$ utilising the GADGET--3 code \citep[see e.g.][]{Springel2005b, Springel2021}, the AGNdT9 variant of the \eagle\ \citep{schaye_eagle_2015, crain_eagle_2015} physics model, and a Planck year 1 cosmology \citep[$\rm \Omega_{M} = 0.307, \; \Omega_{\Lambda} = 0.693, \; $h$ \; = 0.6777$,][]{planck_collaboration_2014}. FLARES adopts an identical
resolution to the fiducial EAGLE simulation, with dark-matter (DM) and gas particle masses of $m_{dm}=9.7\times 10^{6}\ {\rm M_{\odot}}$ 
 and $m_g=1.8\times 10^{6}\ {\rm M_{\odot}}$ respectively, and a softening length of $2.66\ {\rm ckpc}$. The EAGLE model uses a list of 11 elements from \cite{Wiersma2009} to calculate the radiative cooling and photoheating rates. Hydrogen reionisation is implemented as a uniform, time--dependent ionizing background \citep{Haardt2001} that begins at $z$ = 11.5. Finally, star formation is modeled stochastically following the prescription of \cite{DallaVecchia2008,Schaye2004}, coupled with a metallicity-dependent density threshold; and the resulting stellar particles are subject to mass loss and type Ia supernovae.

The 40 FLARES regions were selected from a large (3.2 cGpc)$^3$ low-resolution dark matter only (DMO) simulation \citep{CEAGLE}. The selected regions encompass a wide range of overdensities, $\delta_{14} = -0.4\to 1.0$,  with greater representation at the extremes, particularly extreme over-densities. This enables us to simulate many more massive galaxies than possible using a periodic simulation and the same computational resources. For example, \flares\ contains approximately 100 times as many $M_{\star}>10^{9}\ {\rm M_{\odot}}$ galaxies at $z=10$ than the fiducial \eagle\ reference simulation, despite simulating a comparable total volume. This makes \flares\ ideally suited to studying SMBHs, and particularly AGN, since they are rare and preferentially occur in massive galaxies. This is demonstrated in Figure \ref{fig:sim:nbh}, where we show the total number of SMBHs with $M_{\bullet}/{\rm M_{\odot}}>\{10^{7}, 10^{8}\}$ as a function of redshift for both the (100 Mpc)$^{3}$ EAGLE reference simulation and all FLARES regions combined. At $z=5$ FLARES contains 6 (20) times as many SMBHs with $M_{\bullet}>10^{7} (10^{8})\ {\rm M_{\odot}}$, respectively.  This differences increases with increasing redshift: there are no SMBHs with $M_{\bullet}>10^{8}\ {\rm M_{\odot}}$ in EAGLE at $z>6$, for example, whereas those in FLARES extend to $z=9$. It is important to stress however that while FLARES simulates more massive SMBHs than the EAGLE reference simulation the predicted number densities are consistent once the individual simulations are appropriately weighted (see \S\ref{sec:sim:weighting}, and \S\ref{sec:blackholes:mf}).
In addition to the core suite several regions have been run with model variations. For example, for several regions we ``turned off'' AGN feedback to explore its impact on galaxy properties; this is discussed in \S\ref{sec:impact}.

\subsubsection{Weighting}\label{sec:sim:weighting}

An important consequence of the FLARES strategy is that universal cosmological scaling relations and distribution functions (e.g. the SMBH mass function) cannot be trivially recovered. Instead, it is necessary to weight each simulation/region by how likely it is to occur in the parent DMO simulation. This is described in more detail in \citet{FLARES-I} where we demonstrate its appropriateness by recovering the galaxy stellar mass function.

In short, the $(3.2\ \rm{cGpc})^3$ parent volume is split into grid cells, $2.67\ \rm{cMpc}$ in length. The overdensity in each grid cell is recorded, and the distribution of overdensities is divided into 50 equal-width bins. Each re-simulated region, $r$, is assigned a weight, $w_r$, given by 
\begin{equation}
    w_r = \sum_i f_{r,i} \frac{f_{\rm{parent},\mathit{i}}}{f_{\rm{resims},\mathit{i}}},
\end{equation}
where $f_{r,i}$ is the fraction of grid cells in region $r$ that occupy overdensity bin $i$, $f_{\rm{parent},\mathit{i}}$ is the fraction of grid cells in the complete parent simulation that fall within overdensity bin $i$, and $f_{\rm{resims},\mathit{i}}$ is the fraction of grid cells in all 40 resimulated regions that are contained within overdensity bin $i$. Thus, the weight carried by a region is proportional to how representative that region is of the universal mean.


\subsection{Black Hole Modelling in EAGLE}\label{sec:sim:bhs}

In this section we summarise the SMBH physics utilised by the EAGLE model which is employed by FLARES. For a full description of FLARES see \citet{FLARES-I,FLARES-II}, and for the \eagle\ model see \citet{schaye_eagle_2015} and \citet{crain_eagle_2015}. In short, BHs in the EAGLE model are seeded into sufficiently massive halos and then allowed to grow through accretion and mergers. A fraction of the rest-mass accreted onto the SMBH is able to be radiated away. A fraction of radiated energy is injected into neighbouring gas particles, heating them. 

\subsubsection{Seeding}\label{sec:sim:bhs:seeding}

BHs are seeded into halos exceeding a halo mass of $M_{\rm h} = 10^{10}\ h^{-1} {\rm M_{\odot}}$ by converting the highest density gas particle into a SMBH particle \citep{Springel2005}. BHs carry both a particle mass and a subgrid mass. The particle initial mass is set by that of the converted gas particle while the initial subgrid mass is $M_{\rm \bullet,\ seed} = 10^{5}\ h^{-1} {\rm M_{\odot}}$. The use of a separate subgrid mass is necessary because the black hole seed mass is below the simulation mass resolution. Calculations pertaining to growth and feedback events of the black hole are computed using the subgrid mass, $M_{\bullet}$, while gravitational interactions use the particle mass. When the sub-grid SMBH mass exceeds the particle mass the SMBH particle is allowed to stochastically accrete a neighbouring gas particle. When the sub-grid SMBH mass is much larger than the gas particle mass, the SMBH sub-grid and particle masses effectively grow together. The seed and gas particle masses place an effective lower-limit on the SMBH masses which are robust, or resolved. For this analysis we conservatively assume that BHs are considered resolved at $M_{\bullet}=10^{7}\ {\rm M_{\odot}}$ and focus our analysis on these objects.

\subsubsection{Accretion}\label{sec:sim:bhs:accretion}

The primary growth of SMBHs is through accretion. The EAGLE subgrid accretion model allows the SMBH to accrete material at a maximum rate determined by the Eddington accretion rate scaled by a factor of $1/h$ \citep{McAlpine2020}, i.e.
\begin{equation}
    \dot{M}_{\rm Edd} = \frac{4\uppi G M_{\bullet} m_{\rm p}}{\epsilon _{\rm r} \sigma_{\rm T}c },
    \label{eqn:eddington}
\end{equation}
\noindent
where $G$ is the gravitational constant, $M_{\bullet}$ is the black hole subgrid mass, $m_{\rm p}$ is the proton mass, $\epsilon_{\rm r}$ is the radiative efficiency of the accretion disk, $\sigma_{\rm T}$ is the Thomson cross-section and $c$ is the speed of light. The radiative efficiency in FLARES/EAGLE is set to $\epsilon_{\rm r}=0.1$. 

The floor for the accretion rate is set by,
\begin{equation}
    \dot{M}_{\rm accr} = \dot{M}_{\rm Bondi} \times \mathrm{min} \left ( C_{\rm visc}^{-1}(c_{\rm s}/V_{\upphi})^{3}, 1 \right ) ,
    \label{eqn:accr}
\end{equation}
where $C_{\rm visc}$ is a parameter related to the viscosity of the accretion disc (see next subsection) and $V_{\upphi}$ is the rotation speed of the gas around the SMBH \citep[see equation 16 of][]{Rosas-Guevara2015} and $\dot{M}_{\rm Bondi}$ is the Bondi--Hoyle accretion rate defined as,
\begin{equation}
    \dot{M}_{\rm Bondi} = \frac{4\uppi G^{2} M_{\bullet}^{2} \rho}{(c_{\rm s}^{2} + v^{2})^{3/2}},
    \label{eqn:bondi}
\end{equation}
where $c_{\rm s}$ and $v$ are the speed of sound and the relative velocity of the SMBH and gas, respectively \citep{Bondi1944}.

Finally, this results in the SMBH mass growth rate of
\begin{equation}
    \dot{M}_{\rm \bullet} = (1-\epsilon_{\rm r})\dot{M}_{\rm accr}.
    \label{eqn:bhgrowth}
\end{equation}

The accretion rates of each SMBH are reported at every time--step the SMBH is active and, for all BHs, at the snapshot redshift. Due to numerical noise the rates reported in a single time--step can differ significantly from rates averaged across longer timescales as discussed in \S\ref{sec:blackholes:growth:averaging}.

\subsubsection{Emission}\label{sec:sim:bhs:emission}

The total (bolometric) luminosity radiated by an AGN is simply proportional to the accretion disc or SMBH growth rate as, 
\begin{equation}
    L_{\rm \bullet, bol} = \epsilon_{\rm r}\dot{M}_{\rm accr}c^{2} = \left(\frac{\epsilon_{\rm r}}{1-\epsilon_{\rm r}}\right)\dot{M}_{\bullet}c^{2}.
    \label{eqn:emission:lbol}
\end{equation}
\noindent
As noted previously, the radiative efficiency $\epsilon_{\rm r}$ is assumed to be $0.1$ in the EAGLE model. 


\subsubsection{Mergers}\label{sec:sim:bhs:mergers}

In addition to accretion, SMBHs can also grow by merging. SMBHs are merged if their separation is smaller than three gravitational softening lengths and are within the SMBH smoothing kernel length, $h_{\bullet}$, of each other and have a relative velocity of
\begin{equation}
    v_\mathrm{rel} < \sqrt{G M_{\bullet} / h_{\bullet}},
    \label{eqn:bhvelocity}
\end{equation}
\noindent
where $M_{\bullet}$ is the subgrid mass of the bigger of the merging SMBHs. The limiting velocity given by equation \ref{eqn:bhvelocity} prevents SMBH mergers during the initial stages of galaxy mergers and makes them only possible once some time of the initial merger has passed and the relative velocities have settled. As a consequence galaxies in the EAGLE model can often host multiple black holes. We explore this briefly in the context of FLARES in \S\ref{sec:blackholes:multiplicity}.

\subsubsection{Dynamical Fraction}\label{sec:sim:bhs:repositioning}

Cosmological simulations like EAGLE and FLARES struggle to model the effect of dynamical fraction, since the SMBH particle masses are often comparable in mass, or even less massive, than the surrounding particles \citep[see e.g.][]{Tremmel15}. To account for this, we force SMBHs with $M_{\bullet}<100\ m_{g}$ to be moved towards the minimum of the gravitational potential of the halo they reside in. It has since been shown that this is important to ensure efficient black hole growth, and subsequent feedback \citep{Bahe2022}.
The migration is computed at each simulation time step (given in expansion factor $a$ as $\Delta a = 0.005a$) by finding the location of the particle that has the lowest gravitational potential out of particles neighbouring the SMBH with relative velocities smaller than 0.25$c_{\rm s}$ and distances smaller than three gravitational softening lengths (we use a Plummer--equivalent softening length of 1.8 $h^{-1}$ ckpc). This SMBH migration calculation is crucial in preventing SMBHs in low density, gas poor halos from being stolen by nearby satellite halos \citep{schaye_eagle_2015}. 

\subsubsection{Feedback}\label{sec:sim:bhs:feedback}


AGN feedback in the EAGLE model is modelled with only one feedback channel, contrasting with the multi-mode feedback implemented in other simulations including Simba \citep{Dave2019} and TNG \citep{ILLUSTRIS-Zinger+2020}. In this model, described fully in \citet{schaye_eagle_2015} and based on \citet{Booth2009}, thermal energy is injected stochastically to gas particles neighbouring the SMBH particle, in a kernel--weighted manner. 

The energy injection rate is calculated as a fraction of the total accretion rate, as $\epsilon_{\rm f} \epsilon_{\rm r} \dot{M}_{\rm accr} c^{2}$, where $\epsilon_{\rm f} = 0.15$ is the fraction of the feedback energy that is coupled to the ISM and $\epsilon_{\rm r}=0.1$ is the radiative efficiency of the accretion processes introduced in Eq. \ref{eqn:bhgrowth}. Every time-step energy ($\Delta E=\epsilon_{\rm r} \epsilon_{\rm f} \dot{M}_{\rm accr} c^{2} \Delta t$) is added to reservoir of feedback energy $E_{\bullet}$. If the SMBH has sufficient stored energy to heat at least $n_{\rm heat}$ particles (where here we assume $n_{\rm heat}=1$) of mass $m_{g}$ by $\Delta T_{\rm AGN}$ then the SMBH is allowed to stochastically heat each of its neighbouring particles by increasing their temperature by $\Delta T_{\rm AGN}$. The probability of injecting energy into each nearby gas particle is given by,
\begin{equation}
   P = \frac{E_{\bullet}}{\Delta \epsilon_{\rm AGN}N_\mathrm{ngb} \left \langle m_{\rm g} \right \rangle},
    \label{eqn:energyprobability}
\end{equation}
\noindent
where $\Delta \epsilon_{\rm AGN}$ is the change in internal energy per unit mass of a gas particle corresponding to a temperature increase of $\Delta T_{\rm AGN}$, $N_\mathrm{ngb}$ is the number of gas neighbours of the SMBH and $\left \langle m_{\rm g} \right \rangle$ is their mean mass. The reservoir, $E_{\bullet}$, is then decreased by the amount of the injected energy. 

Since $\Delta T_{\rm AGN}$ directly determines the amount of energy in each AGN feedback event, it is the most important factor in modelling the feedback from SMBH accretion. A larger value results in more energy dumped into the neighbouring particles of the BH, but also makes the feedback events more rare, as the change in internal energy of a gas particle is directly proportional to temperature increase and the probability of energy injection into a gas particle is inversely proportional to the internal energy increase. The temperature increase from AGN feedback was set to a value higher than that from stellar feedback due to gas densities being higher in the vicinity of black holes than they are for typical star-forming gas \citep{crain_eagle_2015}. In FLARES we adopt the approach of the C-\eagle\ simulations \citep{CEAGLE}, using the AGNdT9 subgrid parameter configuration. This configuration is parameterised with $C_\mathrm{visc} = 2\uppi \times 10^{2}$ and $\Delta T_{\rm AGN}=10^{9}\,$K, where the former is a free parameter controlling the sensitivity of the SMBH accretion rate to the angular momentum of the gas and the latter is the temperature increase of gas during AGN feedback.

\subsection{Stellar Emission}

In Section \ref{sec:hosts} we make comparisons between the bolometric luminosities of the SMBHs and the stellar content of galaxies. The spectral energy distribution modelling in FLARES is described in depth in \citet{FLARES-II}; in short, every star particle in each galaxy is associated with an SED based on its mass, age, and metallicity assuming a particular choice of stellar population synthesis model and initial mass function (IMF). In this work we assume \texttt{v2.2.1} of the Binary Population and Spectral Synthesis (BPASS) model \citep{BPASS2.2.1} and a \citet{ChabrierIMF} IMF. While in \citet{FLARES-II} we also consider the impact of reprocessing by dust and gas here we ignore those effects since we are only interested in the bolometric output.

\section{Blackhole Properties}\label{sec:blackholes}

In this section we explore the physical properties of SMBHs and their host galaxies in FLARES. As previously noted (in \S\ref{sec:sim:bhs}), a consequence of the simulation resolution and modelling choices (i.e. SMBH seed mass) is that we can only be confident in the properties of SMBHs with masses $M_{\bullet}>10^{7}\ {\rm M_{\odot}}$, which we consider ``resolved''. Consequently, we focus our attention on these systems though we do explore predictions at lower-masses.

\subsection{Multiplicity}\label{sec:blackholes:multiplicity}

As noted in \S\ref{sec:sim:bhs:mergers}, galaxies in the EAGLE model can host multiple black holes, though many of these will eventually merge. We find that while most massive galaxies simulated by FLARES contain multiple black holes, in the vast majority ($\approx 94\%$) of galaxies hosting at least one resolved SMBH, the most-massive SMBH accounts for $>90\%$ of the total mass. Out of the galaxies hosting a resolved SMBH, around 2\% of them host multiple resolved black holes. 

\subsection{SMBH density and mass Function}\label{sec:blackholes:mf}

\begin{figure}
    \centering
    \includegraphics[width=\columnwidth]{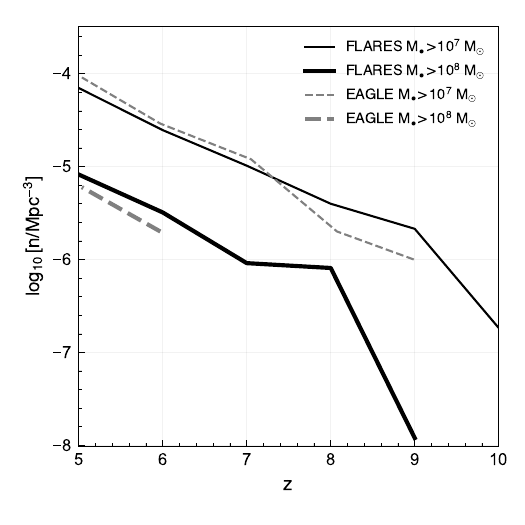}
    \caption{The evolution of the SMBH number density predicted by EAGLE and FLARES at $z=10-5$ for $\Mbh>10^7\ \Msun$ and $\Mbh >10^8\ \Msun$.}
    \label{fig:physical:Mbh_n}
\end{figure}

Next, we explore the evolution of the SMBH number density, from $z=10\to 5$, in Figure \ref{fig:physical:Mbh_n}. As expected we find good agreement between FLARES and EAGLE, validating our simulation strategy and weighting scheme. 

The density drops by $\approx 10\times$ between $z=5\to 7$, irrespective of the mass threshold. This is comparable to the drop in density of galaxies with $M_{\star}>10^{10}\ {\rm M_{\odot}}$ \citep[see][]{FLARES-I}, but faster than the drop in lower-mass galaxies. This evolution appears to continue at $z>7$ but here the numbers simulated are small, leading to a large statistical uncertainty. 

\begin{figure*}
    \centering
    \includegraphics[width=2\columnwidth]{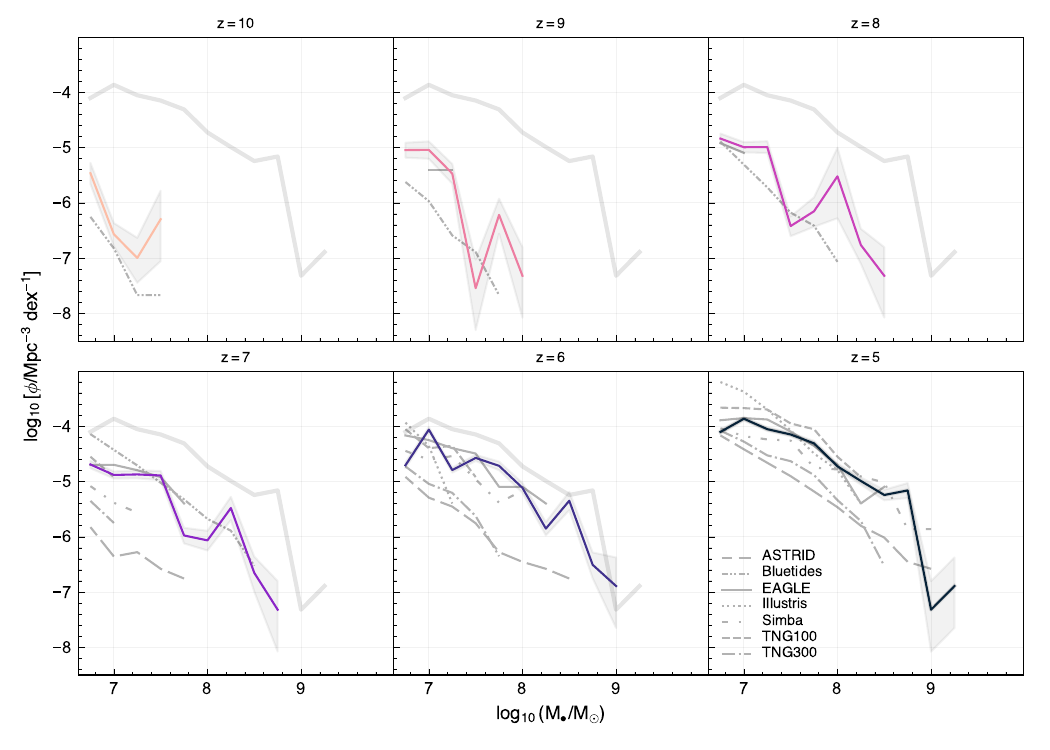}
    \caption{The evolution of the SMBH mass function predicted by \flares\ at $5\le z\le 10$. The solid coloured lines in both panels show the total mass function (i.e. for all SMBHs). The shaded region around this provides an estimate of the statistical uncertainty using the $68\%$ Poisson confidence interval. Also shown on each panel are models predictions from ASTRID, Bluetides, EAGLE, Illustris, Simba, TNG100, and TNG300.}
    \label{fig:physical:Mbh_DF}
\end{figure*}

Figure \ref{fig:physical:Mbh_DF} shows the SMBH mass function predicted by FLARES, again from $z=10\to 5$ highlighting its evolution. Figure \ref{fig:physical:Mbh_DF} also includes an estimate of the statistical uncertainty in each bin based on $68\%$ Poisson confidence interval. However, this is the minimum possible uncertainty based simply on the \emph{total} number of SMBHs in each mass bin. In the context of FLARES this is likely a significant underestimate since mean-density simulation have small numbers of SMBHs (and thus individually large statistical uncertainties) but high weighting. The mass function also drops by an order of magnitude from $M_{\bullet}=10^{7}\to 10^{8}\ {\rm M_{\odot}}$, largely independent of redshift. There is tentative evidence of a steeper drop at $M_{\bullet}>10^{8}\ {\rm M_{\odot}}$, but this is complicated by the small sample size and thus large statistical uncertainty. The total mass function shown in Figure \ref{fig:physical:Mbh_DF} is provided in Table \ref{tab:Mbh_DF} and electronically.

\begin{table}
\caption{The SMBH mass function at $5\le z\le 10$ predicted by FLARES. An electronic version of this is available at: \url{https://github.com/stephenmwilkins/flares_agn_data}.}\label{tab:Mbh_DF}
\begin{tabular}{lcccccc}
\hline
 & \multicolumn{6}{c}{$\log_{10}(\phi/{\rm Mpc^{-3}\ dex^{-1}})$} \\
$\log_{10}(M_{\bullet}/{\rm M_{\odot}})$ & $z=10$ & $z=9$ & $z=8$ & $z=7$ & $z=6$ & $z=5$ \\
\hline
7.25 & -7.0 & -5.47 & -4.99 & -4.87 & -4.79 & -4.05 \\
7.5 & -6.3 & -7.54 & -6.42 & -4.89 & -4.57 & -4.15 \\
7.75 & -  & -6.22 & -6.15 & -5.98 & -4.72 & -4.31 \\
8.0 & -  & -7.32 & -5.52 & -6.06 & -5.11 & -4.73 \\
8.25 & -  & -  & -6.76 & -5.48 & -5.85 & -4.99 \\
8.5 & -  & -  & -7.32 & -6.65 & -5.35 & -5.24 \\
8.75 & -  & -  & -  & -7.32 & -6.51 & -5.16 \\
9.0 & -  & -  & -  & -  & -6.89 & -7.32 \\
9.25 & -  & -  & -  & -  & -  & -6.89 \\
\hline
\end{tabular}
\end{table}

\subsubsection{Observational Constraints}\label{sec:blackholes:mf:observations}

As a key physical distribution function the SMBH mass function has been the focus of considerable observational study. However, it is very challenging to measure accurately. First, it is necessary to constrain individual SMBH masses. Assuming the broad-line region (BLR) is virialised, the SMBH mass can be estimated from the motion of the BLR and its radius. Based on reverberation mapping of local AGN it has been established that there is a tight correlation between the size of the emitting region and the continuum luminosity \citep{Kaspi2000}. Observations of a line-width and continuum luminosity can then be used to constrain the masses of SMBH. While originally established for the H$\beta$ line, this technique has been extended to other broad emission lines. SMBH mass functions measured in this way are more accurately described as \emph{broad line} SMBH mass functions and, since SMBH masses do not only scale with luminosity, can have a complex completeness function making them difficult to compare with simulation predictions. Some observational studies also attempt to infer \emph{total} SMBH mass function by correcting the \emph{broad line} SMBH mass function to include both obscured (type 2) AGN and inactive SMBHs (i.e. SMBHs with luminosities below the sensitivity of the particular survey). An integral part of this requires inferring the bolometric luminosity of the SMBH which can involve uncertain bolometric corrections.

\citet{He2023} recently combined SDSS observations with a fainter Hyper Suprime-Cam selected sample with spectroscopic follow-up to study the SMBH mass function at $z\approx 4$. This samples ranges from $M_{\bullet}=10^{7.5-10.5}\ {\rm M_{\odot}}$ and bolometric luminosities $10^{45.5-47.5}\ {\rm erg\ s^{-1}}$. These constraints are consistent with our total mass function function at $\approx 10^{9}\ {\rm M_{\odot}}$ but fall significantly below our predictions at lower-masses. However, the \citet{He2023} mass function is measured at $z=4$, not at $z=5$ where the FLARES predictions lie. Extrapolating the FLARES $z=5$ SMBH mass function to $z=4$, based on the previous evolution, would suggest a density increase of around 0.5 dex. Secondly, the \citet{He2023} mass function only includes unobscured (i.e. type 1) and active SMBH and thus provides only a lower-limit on the true mass function. However, \citet{He2023} also attempt to constrain the \emph{total} mass function, making corrections for obscuration and in-active SMBHs. The density of $M_{\bullet}=10^{8-9}\ {\rm M_{\odot}}$ BHs is $\sim 10-100\times$ larger than density of active broad line SMBHs, with the range sensitive to the modelling assumptions. The upper-end of this correction would elevate the total mass function above our predicted mass function across the full mass-range. 

\emph{JWST} is now enabling a similar approach at higher-redshift and lower luminosities. Several studies \citep{Harikane2023, Matthee23, Maiolino2023a} have already employed this method to infer SMBH masses, and in the case of \citet{Matthee23} the \emph{broad-line} AGN SMBH mass function. The \emph{raw} observational constraints of \citet{Matthee23} provide a good match to our predictions for \emph{all} SMBH. However, for the same reasons described above, the \citet{Matthee23} mass function will provide a lower-limit on the total mass function suggesting that the \emph{true} (completeness corrected) mass function will lie above our predictions. However, this also suggests possible tension between \citet{Matthee23} and \citet{He2023}. 

An alternative observational test is to compare against the bolometric luminosity function since this, in principle, will suffer \emph{fewer} completeness issues. However, this has its own complications. This is discussed in \S\ref{sec:blackholes:growth:lf} and we defer a discussion of potential modelling changes until there.


\subsubsection{Model comparisons}\label{sec:blackholes:mf:models}

Figure \ref{fig:physical:Mbh_DF} also shows predictions from several other cosmological hydrodynamical simulations including ASTRID, Bluetides, EAGLE, Illustris, Simba, TNG100, and TNG300. This Figure immediately demonstrates the power of the FLARES approach - only the very-large $(400/h\ {\rm Mpc})^{3}$ volume Bluetides simulation mass function extends as far as FLARES, and only then to $z=7$ where the simulation stopped. For all the other simulations FLARES significantly extends the mass range probed. As expected, the FLARES predictions closely match the EAGLE predictions (where they overlap) but are also similar (i.e. $<0.5$ dex difference) to Illustris, Simba, TNG100, and Bluetides (at $z=7$, less so at $z>7$). The agreement with ASTRID and TNG300 is significantly weaker with both predicting $\sim 1$ dex fewer massive SMBHs than FLARES at $z=5$ with the disagreement widening at higher-redshift.

The differences seen here reflect the different ways in which the seeding, dynamics, growth, and feedback are modelled by the different simulations. A thorough comparison between many of these models is presented in \citet{Habouzit2022a} at low-redshift and \citet{Habouzit2022b} at high-redshift. However, unfortunately, since each model\footnote{Except FLARES and EAGLE, since FLARES is based on a variant of the EAGLE model.} has several differences in both the SMBH and wider physics it is impossible to decisively identify the source of differences. A curiosity however is the large difference between ASTRID and FLARES/EAGLE; ASTRID seeds lower-mass halos, using a comparable mass seed\footnote{ASTRID assumes a probabilistic approach for the seed masses, drawing from a power-law distribution $M_{\bullet, {\rm seed}}/h^{-1}\ {\rm M_{\odot}}=[3\times 10^{4}, 3\times 10^{5}]$ with power-law index $n=-1$.}, and permits more rapid accretion but results in $\sim 10\times$ fewer massive BHs at $z=5$. This suggests that the AGN feedback, or some other feature of the model, is key here. A priority for the next phase of FLARES is to systemically explore the impact of the SMBH modelling choices, including the choice of parameters. This should provide a much clearer understanding of the key factors driving the growth of SMBHs in the early Universe.

\subsubsection{Environmental Dependence}\label{sec:blackholes:mf:environment}

\begin{figure}
    \centering
    \includegraphics[width=1\columnwidth]{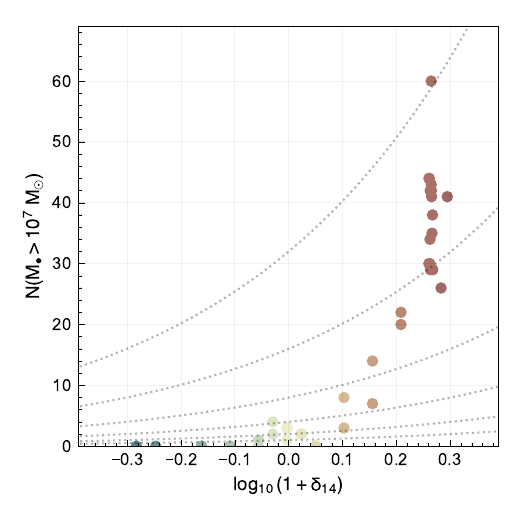}
    \includegraphics[width=1\columnwidth]{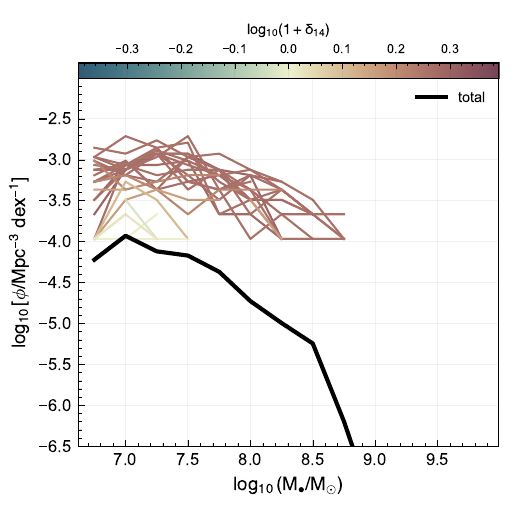}
    \caption{Environmental dependence of the number (top panel) and mass function (bottom panel) of SMBHs predicted by FLARES at $z=5$. The top panel shows the number of robust $M_{\bullet}>10^{7}\ {\rm M_{\odot}}$ SMBHs in each simulated region as a function of the region over-density $\delta_{14}$. Dotted lines indicate the behaviour if $N\propto \delta + 1$. The bottom panel shows the SMBH mass function for every individual simulation with at least one SMBH in the mass range}, colour coded by the region over-density $\delta_{14}$. Also shown here in black is the composite SMBH mass function found by combining all forty simulations with the appropriate weighting. The horizontal grey line denotes the density corresponding to a single object in an individual FLARES simulation. 
    \label{fig:physical:environment}
\end{figure}

One of the strengths of FLARES is its ability to probe the effect of environment on galaxy formation at high-redshift. In Figure \ref{fig:physical:environment} we show how the total number (top-panel) and mass function (bottom panel) of SMBHs varies between simulations and thus environment, $\log_{10}(1+\delta_{14})\approx -0.3\to 0.3$. 

These figures reveal, unsurprisingly, that SMBHs in FLARES are significantly biased. Virtually all of the resolved ($>10^{7}\ {\rm M_{\odot}}$) SMBHs simulated in FLARES are found in the most extreme regions. While there are $\approx 700$ $>10^{7}\ {\rm M_{\odot}}$ SMBHs across the FLARES simulations at $z=5$, only $\approx 10$ are in regions with $\delta_{14}\le 0.0$. This bias is even more extreme for more massive SMBHs, with SMBHs of $>10^{8}\ {\rm M_{\odot}}$ only lying in the most extreme over-densities that we simulate. The SMBH mass function of our most over-dense regions ($\log_{10}(1+\delta_{14}\approx 0.25$) is around $1$ (1.5) ${\rm dex}$ higher at $M_{\bullet}=10^{7}\ {\rm M_{\odot}}$ ( $M_{\bullet}=10^{8}\ {\rm M_{\odot}}$) higher than the weighted average. 

This then suggests that massive SMBHs (and bright AGN) should pinpoint over-dense regions. This has recently been explored by \citet{Eilers2024} who used JWST/NIRCam observations to explore the environments of 4 luminous ($\sim 10^{46}\ {\rm erg\ s^{-1}})$ quasars at $z\ge 6$. \citet{Eilers2024} found that these quasars do indeed on average trace large over-densities but they do not necessarily trace the rarest and highest density peaks. However, it is not immediately straightforward to compare the \citet{Eilers2024} results with our predictions. This is because the \citet{Eilers2024} quasars are brighter than any AGN in FLARES at $z\ge 6$, they use a smaller search radius, and use [O\textsc{iii}] emitting galaxies.

\subsection{Growth}\label{sec:blackholes:growth}

As described in \S\ref{sec:sim:bhs} SMBHs in the EAGLE model grow through both accretion (\S\ref{sec:sim:bhs:accretion}) and mergers (\S\ref{sec:sim:bhs:mergers}), albeit with mergers producing only a small fraction of the overall growth of resolved SMBHs in FLARES. We defer an exploration of SMBH mergers in FLARES to a work in preparation (Liao et al. \emph{in-prep}) and focus here on growth through accretion, including making predictions for the accretion rates and bolometric luminosities of SMBHs in FLARES.

\subsubsection{Accretion rate definitions}\label{sec:blackholes:growth:averaging}

By default the accretion rate associated with each SMBH in a particular snapshot is that calculated in the most recent time-step. However, these \emph{instantaneous} accretion rates are numerically noisy. Fortunately, within FLARES SMBH accretion rates are recorded at every time-step, permitting an exploration of accretion rate histories.

\begin{figure}
    \centering
    \includegraphics[width=\columnwidth]{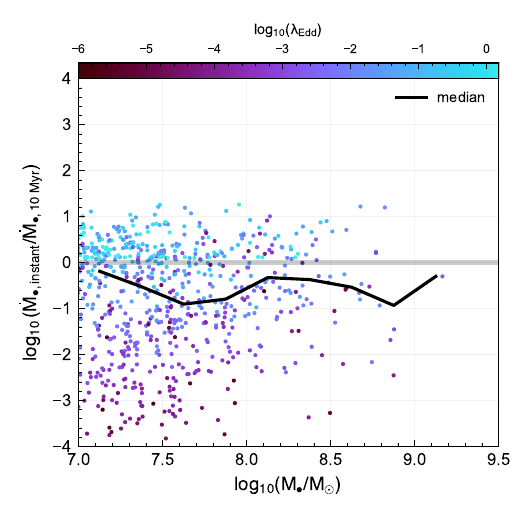}
    \caption{The ratio of the instantaneous accretion rate to the accretion rate averaged over the preceding 10 Myr for SMBHs at $z=5$. The solid line denoted the median in $M_{\bullet}$ bins. Individual BHs are colour coded by their Eddington ratio ($\lambda$) based on the instantaneous accretion rate.}
    \label{fig:blackhole_mass-accretion_rate-averaging}
\end{figure}

As an alternative to the instantaneous accretion rate we can average the accretion over a longer timescale ( $t_{\rm avg}/{\rm Myr}=10-200$) in an attempt to reduce the numerical noise. Figure \ref{fig:blackhole_mass-accretion_rate-averaging} shows the ratio of the instantaneous accretion rate to the accretion rate averaged over the preceding 10 Myr for SMBHs at $z=5$. This reveals that the majority of galaxies have average accretion rates larger than their instantaneous rate. However, the total accretion integrated over all galaxies is similar as would be expected for a large number of SMBHs.

\begin{figure}
    \centering
    \includegraphics[width=\columnwidth]{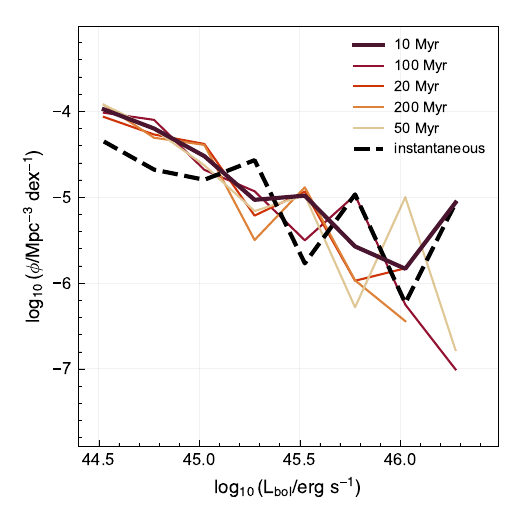}
    \caption{The bolometric luminosity function of SMBHs using both the instantaneous accretion rates (thick black line) and varying averaging timescales $\in\{10, 20, 50, 100, 200\}\ {\rm Myr}$.}
    \label{fig:bolometric_luminosity_function-averaging}
\end{figure}

While the total accretion integrated across all SMBHs is similar irrespective of the timescale there is an impact on the shape of the accretion rate distribution function, or bolometric luminosity function. Figure \ref{fig:bolometric_luminosity_function-averaging} shows the bolometric luminosity function calculated assuming the different accretion timescales $\in\{10, 20, 50, 100, 200\}\ {\rm Myr}$. While the density of very-luminous ($>10^{46}\ {\rm erg\ s^{-1}}$) SMBHs remains relatively unchanged, the number of lower luminosity, but still bright ($<10^{45}\ {\rm erg\ s^{-1}}$) SMBHs increases. This is due to many SMBHs with very low instantaneous accretion rates having significantly higher averaged rates as demonstrated by Figure \ref{fig:blackhole_mass-accretion_rate-averaging}.

Since the instantaneous accretion rates are subject to numerical noise we decide to adopt the 10 Myr averaged accretion rate as our fiducial measure in the remainder of our analysis. It is important to stress however that the luminosities of real AGN are well established to vary on timescales much shorter than this, or possible to resolve in any cosmological simulation. This is a clear limitation of this type of analysis.

\subsubsection{Correlation with SMBH mass}\label{sec:blackholes:growth:mass}

\begin{figure}
    \centering
    \includegraphics[width=\columnwidth]{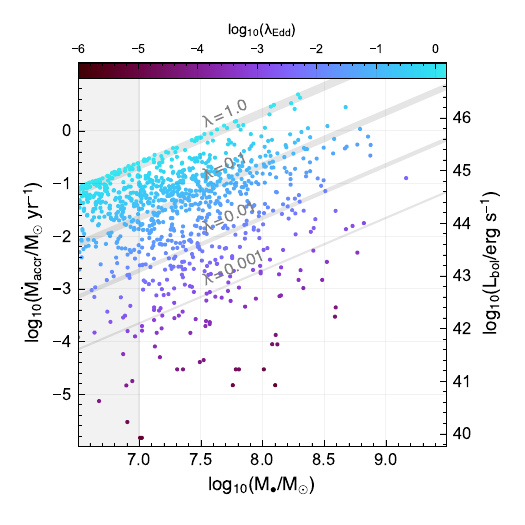}
    \includegraphics[width=\columnwidth]{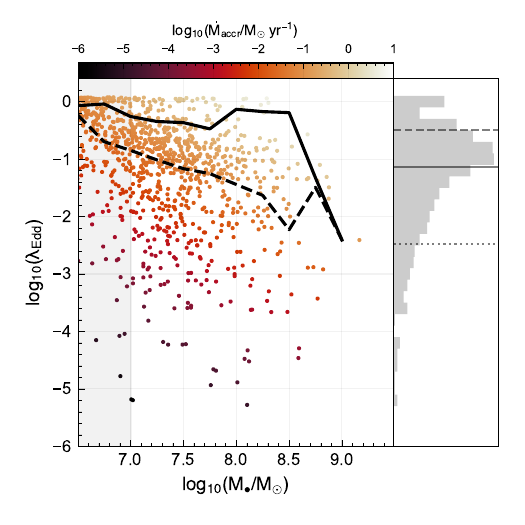}
    \caption{\emph{Top:} The relationship between SMBH mass and accretion disc accretion rate (left-axis) and bolometric luminosity (right-axis) predicted by FLARES at $z=5$. Diagonal lines denote fixed Eddington ratios $\lambda\in\{0.001, 0.01, 0.1, 1.0\}$ and objects are colour-coded by their Eddington ratio. \emph{Bottom:} The relationship between SMBH mass and Eddington ratio $\lambda$. The thick dashed line denotes the binned median while the solid line denotes the binned median weighted by the accretion rate. The right-hand panel shows the normalised distribution of Eddington ratios for $M_{\bullet}>10^{7}\ {\rm M_{\odot}}$. The dotted, solid, and dashed lines denote the 15.8th, 50th, and 84.2th percentiles respectively.}
    \label{fig:physical:mbh_Mbhdot}
\end{figure}

We next explore, in Figure \ref{fig:physical:mbh_Mbhdot}, predictions for the relationship between SMBH mass ($M_{\bullet}$) and the accretion rate, bolometric luminosity (top--panel) and Eddington ratio $\lambda_{\rm Edd}$ (bottom--panel). As noted previously we can only be confident in our predictions of SMBHs with $M_{\bullet}>10^{7}\ {\rm M_{\odot}}$. Since a large fraction of SMBHs are accreting at the Eddington limit this suggests our predictions for the bolometric luminosities of SMBHs are, conservatively, complete above the Eddington luminosity of a $10^{7}\ {\rm M_{\odot}}$ SMBH, i.e. $\approx 10^{45}\ {\rm erg/s}$ ($\approx 2.5\times 10^{11}\ {\rm L_{\odot}}$). 

SMBHs in FLARES exhibit a wide range of accretion rates at fixed mass, extending close to the imposed maximum ($1/h$ $\times$ the Eddington rate). Few SMBHs have accretion rates at the maximum since (as discussed in \S\ref{sec:blackholes:growth:averaging}) we average over the last 10 Myr instead of using instantaneous accretion rates. For resolved SMBHs the median Eddington ratio is around $0.1$ below which it gradually drops such that there are very few galaxies with $\lambda<10^{-4}$. While the number of SMBHs with ratios higher than the median also drops, they ``pile-up'' at towards the imposed limit. The binned median Eddington ratio (denoted by the dashed line in the bottom--panel of Figure \ref{fig:physical:mbh_Mbhdot}) drops by around $1.0$ dex $M_{\bullet}>10^{7}-10^{9}\ {\rm M_{\odot}}$. Figure \ref{fig:physical:mbh_Mbhdot} also shows the binned median Eddington rate but weighted by the accretion rate (solid). As would be expected this is biased towards higher accretion rates, except at the highest masses where the small numbers lead to convergence. Thus, while the typical Eddington ratios are $\sim 0.1$ most accretion (and thus most energy is produced) is SMBHs with much higher Eddington ratios ($\sim 1$).

Another feature to note is that the most luminous SMBHs in FLARES are not necessarily the most massive. While more massive SMBHs on average have higher luminosities, the mass function is so steep that there are many more lower-mass SMBHs resulting in them making up a larger share of the most luminous SMBHs. For example, at $z=5$ the most luminous ($L_{\rm bol}\approx 10^{47}\ {\rm erg\ s^{-1}}$) SMBH has a mass of $M_{\bullet}\approx 2\times 10^{8}\ {\rm \Msun}$ compared to the most massive SMBH which has $M_{\bullet}> 10^{9}\ {\rm \Msun}$. 

Recent observational constraints from \emph{JWST} suggest Eddington ratios of $\lambda = 0.01-1.0$ in $\Mbh>10^{7}\ \Msun$ \citep[e.g.][]{Maiolino2023a}. Our predictions extend to lower values of $\lambda$ but this may simply reflect an observational bias, since SMBHs with lower ratios will be less luminous and thus possibly missed. Thus, at present, there does not appear to be a contradiction between our predictions and the observations of these properties.

\subsubsection{Bolometric Luminosity Function}\label{sec:blackholes:growth:lf}

\begin{figure*}
    \centering
    \includegraphics[width=2.\columnwidth]{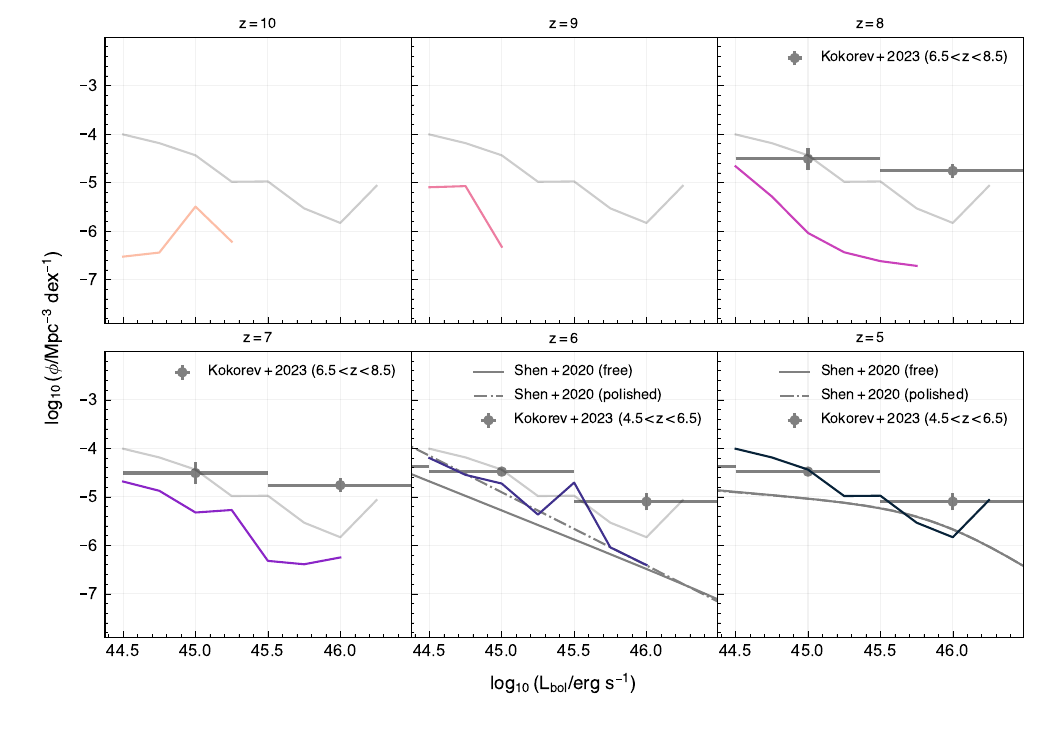}
    \caption{Bolometric luminosity function for SMBHs at $z\in[5, 10]$ predicted by FLARES. Also shown are high-redshift predictions for the SMBH bolometric luminosity function from the \citet{Shen2020} empirical models of  and recent observational constraints on the SMBH bolometric luminosity function from \citet{Kokorev2024}. The faint grey line in each panel shows the $z=5$ prediction.}
    \label{fig:photometric:bol:lf}
\end{figure*}

\begin{table*}
\centering
\caption{The SMBH bolometric luminosity function at $5\le z\le 10$ predicted by FLARES. An electronic version of this is available at: \url{https://github.com/stephenmwilkins/flares_agn_data}.}\label{tab:Lbol_DF}
\begin{tabular}{lcccccc}
\hline
 & \multicolumn{6}{c}{$\log_{10}(\phi/{\rm Mpc^{-3}\ dex^{-1}})$} \\
$\log_{10}(L_{\rm bol}/{\rm erg\ s^{-1}})$ & $z=10$ & $z=9$ & $z=8$ & $z=7$ & $z=6$ & $z=5$ \\
\hline
44.5 & -6.53 & -5.1 & -4.67 & -4.68 & -4.2 & -4.01 \\
44.75 & -6.44 & -5.07 & -5.29 & -4.88 & -4.55 & -4.19 \\
45.0 & -5.5 & -6.32 & -6.04 & -5.32 & -4.73 & -4.44 \\
45.25 & -6.22 & - & -6.44 & -5.27 & -5.37 & -4.98 \\
45.5 & - & - & -6.62 & -6.32 & -4.71 & -4.98 \\
45.75 & - & - & -6.72 & -6.39 & -6.04 & -5.53 \\
46.0 & - & - & - & -6.25 & -6.41 & -5.83 \\
46.25 & - & - & - & - & - & -5.07 \\
\end{tabular}
\end{table*}

As noted in \S\ref{sec:sim:bhs:emission}, since we assume a fixed radiative efficiency (in our case $\epsilon_{\rm r} = 0.1$) the bolometric luminosity simply scales with the accretion disc accretion rate, allowing us to explore them interchangeably. Since the bolometric luminosity is observationally accessible in Figure \ref{fig:photometric:bol:lf}, we explore predictions for the SMBH bolometric luminosity function. Due to our conservative completeness limit there are only a relatively small number of objects with luminosities above this limit, resulting in noisier predictions than the SMBH mass function. Nevertheless, this analysis reveals a clear evolution from $z=10\to 5$ with the density of SMBHs with $L_{\rm bol}>10^{45}\ {\rm erg\ s^{-1}}$ increasing by around a factor $100$.

In Figure \ref{fig:photometric:bol:lf}, we also compare our predictions to two recent observational studies. First we compare against the ``free'' and ``polished'' variants of the quasar bolometric luminosity function presented in \citet{Shen2020} at $z=5$ and $z=6$.  These luminosity functions are derived by converting multiple observations of monochromatic quasar luminosity functions in the UV, optical, hard X-ray, and mid-IR into bolometric luminosity functions and finding the best fit. In the ``free'' variant the fitting of $\phi_{\star}$ is left free at each redshift. However, at high-redshift there is considerable observational uncertainty. In the ``polished'' variant $\phi_{\star}(z)$ is assumed to be linear with observations at $z=0.4-3.0$ used to define the relationship between $z$ and $\phi_{\star}$. At $z=5$ the normalisation at the bright-end ($L\sim 10^{45.5-46}\ {\rm erg\ s^{-1}}$) is in good agreement with the \citet{Shen2020} model; however, at fainter luminosities FLARES predicts more SMBHs. At $z=6$ FLARES predicts a similar shape and normalisation to the ``polished'' variant across the full luminosity range. However, it is important to note that the FLARES predictions cover the low-luminosity limit of the \citet{Shen2020} model.

As noted in the introduction, \emph{JWST} has recently begun placing observational constraints on the demographics and properties of AGN at high-redshift. In the context of the bolometric luminosity function, this has recently been constrained by observations of the H$\alpha$ recombination line luminosity \citep{Greene23} and UV luminosities of photometrically identified AGN \citep{Kokorev2024}. The bolometric luminosity function constraints of \citet{Kokorev2024} at $4.5<z<6.5$ (shown in the the $z=5$ and $z=6$ panels) and $6.5<z<8.5$ (shown in the the $z=7$ and $z=8$ panels) are shown in Figure \ref{fig:photometric:bol:lf}\footnote{Due to an analysis error, the number densities presented in v1 of \citet{Kokorev2024} were too high by $\approx 0.2$ dex. Here we use number densities to appear in an updated version of the manuscript.}. The \citet{Greene23} constraints at $4.5<z<6.5$ are very similar to the \citet{Kokorev2024} constraints and we omit them for clarity here. Our constraints at $z=5$ are approximately consistent with the \citet{Kokorev2024} $4.5<z<6.5$ constraints. However, at higher redshift our predictions diverge from the \citet{Kokorev2024} observations with the predicted bolometric luminosity function falling off much faster than the observations.

The cause of this tension could lie with the observations, the model, or a combination. Observationally, it is first worth noting that existing samples are small and susceptible to cosmic variance. As \emph{JWST} and later \emph{Euclid} survey larger areas sample sizes will increase and cosmic variance will be mitigated. Second, there could be significant contamination from the SMBH's host galaxy. Indeed, as we will see in \S\ref{sec:hosts:bolometric_relative} the luminosities probed here are in the regime where the host and SMBH provide similar contributions to the bolometric luminosity. In addition, observational constraints on bolometric luminosities at high-redshift currently come from observations encompassing only a small part of the spectrum, either the rest-frame UV or the H$\alpha$ line. Inferring the bolometric luminosity then requires the assumption of a bolometric correction, often based on low-redshift observations. However, bolometric corrections are predicted to depend on the properties (including mass, accretion rate etc.) of the SMBH \citep[e.g.][]{KD18}; assuming a single correction will then introduce noise and potentially bias. This can be overcome either by expanding the wavelength range of observations (i.e. capturing a larger fraction of the bolometric emission) or by forwarding modelling the simulated blackholes to predict the observed emission. This is a current focus of work with results expected in a companion work (Wilkins et al. \emph{in-prep}).

There are also several modelling factors which may explain  this tension and also possible the \emph{potential} tension in the mass function. 

First, as noted in \S\ref{sec:blackholes:growth:averaging}, we average accretion rates over a 10 Myr timescale to reduce numerical noise. However, the emission from SMBHs is well established to often vary on much shorter timescales. While this would impact the shape of the bolometric luminosity it seems extremely unlikely that it could account for the significant tension at the highest redshifts. 

It's possible that these tensions could also be explained by changes to the model. Properly evaluating the impact of changes to the model is beyond the scope of this work, since any change will manifest in complex ways, impacting the SMBH mass and luminosity function, as well as the properties of host galaxies. For example, a boost to the accretion at one time may lead to a subsequent suppression due to feedback. This type of exploration is a key priority for the next phase of the FLARES project. 

However, it is nevertheless useful to consider the impact of some simple changes. First, we could assume a larger radiative efficiency $\epsilon_{\rm r}$. At fixed accretion rate this would boost the bolometric luminosity; however, it would, assuming the same Eddington limiter, also reduce SMBH growth through the impact on Eddington rate (Equation \ref{eqn:eddington}) and SMBH mass growth (Equation \ref{eqn:bhgrowth}). Since the Bondi accretion rate is proportional to $M_{\bullet}^2$ such a change would likely suppress the luminosities of the most luminous SMBHs further exasperating the tension. Alternatively, we could relax the Eddington limiter. Since the most luminous SMBHs are growing at (or close to) the limit this would likely boost the growth and luminosities of the most luminous SMBHs. While this may create better agreement at the highest redshifts, it would likely create a new tension at lower redshift ($z\approx 5$) where the agreement is currently good. Another option is to increase the seed mass; this would likely manifest in larger masses and accretion later one; again removing tension at the highest redshifts, albeit at the expense of adding tension at lower-redshift.


\section{Relation to Host Galaxy Physical Properties}\label{sec:hosts}

We now explore the correlations between SMBH properties and their hosts. In common with the other FLARES analyses, the properties (e.g. the stellar mass - $M_{\star}$) are based on all star particles, associated with the bound sub-halo, within a 30 kpc aperture centred on the potential minimum of the sub-halo.

\subsection{$M_{\bullet}-M_{\star}$ scaling relationship}\label{sec:hosts:mbh-mstar}

\begin{figure}
    \centering
    \includegraphics[width=.9\columnwidth]{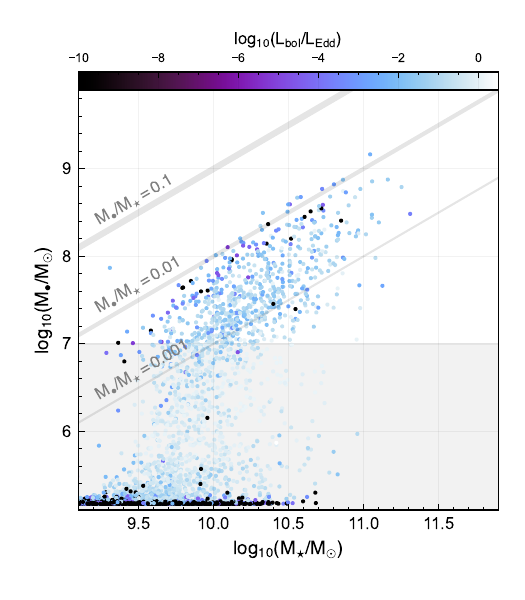}
    \includegraphics[width=.9\columnwidth]{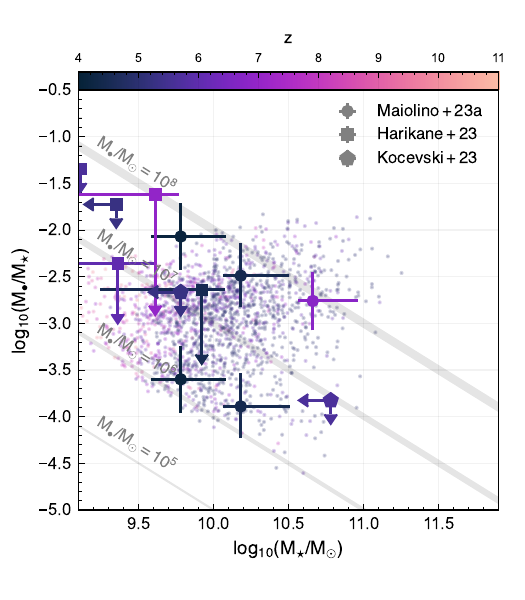}
    \caption{The relationship between stellar mass ($M_{\star}$) and SMBH mass ($M_{\bullet}$) predicted by FLARES for galaxies at $z=5$. In the top-panel only FLARES predictions are shown with individual objects colour-coded by their Eddington ratio $\lambda$. The diagonal lines denote fixed values of $M_{\bullet}/M_{\star}$. The lower-panel instead shows $M_{\bullet}/M_{\star}$ across $5<z<10$, but restricts to objects with $L_{\rm bol}>10^{44}\ {\rm eg\ s^{-1}}$, and also includes observations at $z\sim 5$ from \citet{Maiolino2023a}, \citet{Harikane2023} and \citet{Kocevski23}.}
    \label{fig:physical:Mstar_Mbh_Mbhdot}
\end{figure}

We begin by exploring the correlation between SMBH mass and the stellar content of their host galaxies. The top-panel of Figure \ref{fig:physical:Mstar_Mbh_Mbhdot} shows the relationship between $M_{\bullet}$ and $M_{\star}$ predicted for galaxies at $z=5$. For SMBHs with masses $>10^{7}\ {\rm M_{\odot}}$, the ratio of stellar to SMBH mass ($M_{\star}/M_{\bullet})$ is mostly in the range $100-2000$. There exist only a small number of galaxies where the SMBH has grown to exceed 1 per cent of the stellar mass. However, there are a significant number of galaxies in which the SMBH has yet to grow significantly beyond the seed mass, including relatively massive galaxies (i.e. those with $>10^{10}\ {\rm M_{\odot}}$). Consequently, in FLARES a tight relationship has yet to develop by $z=5$.

\subsubsection{Comparison with Observational Constraints}


With the advent of \emph{JWST}, measurements of both SMBH and stellar masses have become possible to redshifts and luminosities probed by FLARES. Several studies \citep[e.g.][]{Kocevski23, Harikane2023, Maiolino2023a} have now measured stellar and SMBH masses of galaxies at $z>5$, albeit with small samples. These observations are included in the lower-panel of Figure \ref{fig:physical:Mstar_Mbh_Mbhdot} where we limit the FLARES predictions to $L_{\rm bol}>10^{44}\ {\rm eg\ s^{-1}}$ to better align with the completeness of the observations (and also present $M_{\bullet}/M_{\star}$ instead of $M_{\bullet}$ on the $y$-axis). This reveals an overall good correspondence with the majority of observational constraints intersecting with the FLARES predictions. However, a small number of the \citet{Maiolino2023a} observations extend to mass ratios beyond those predicted by FLARES at low stellar masses. While the inferred properties of these objects is subject to large uncertainty this may be hinting at a further potential tension with the model. Possibilities here include the resolution of the simulation but also the choice of halo-seeding mass, seed mass, the limitations on growth, and even the feedback. Like the other possible tensions it is clear that further observations are required.

\subsection{Relative Contribution of SMBHs to the Bolometric Luminosities of Galaxies}\label{sec:hosts:bolometric_relative}

\begin{figure*}
    \centering
    \includegraphics[width=2.0\columnwidth]{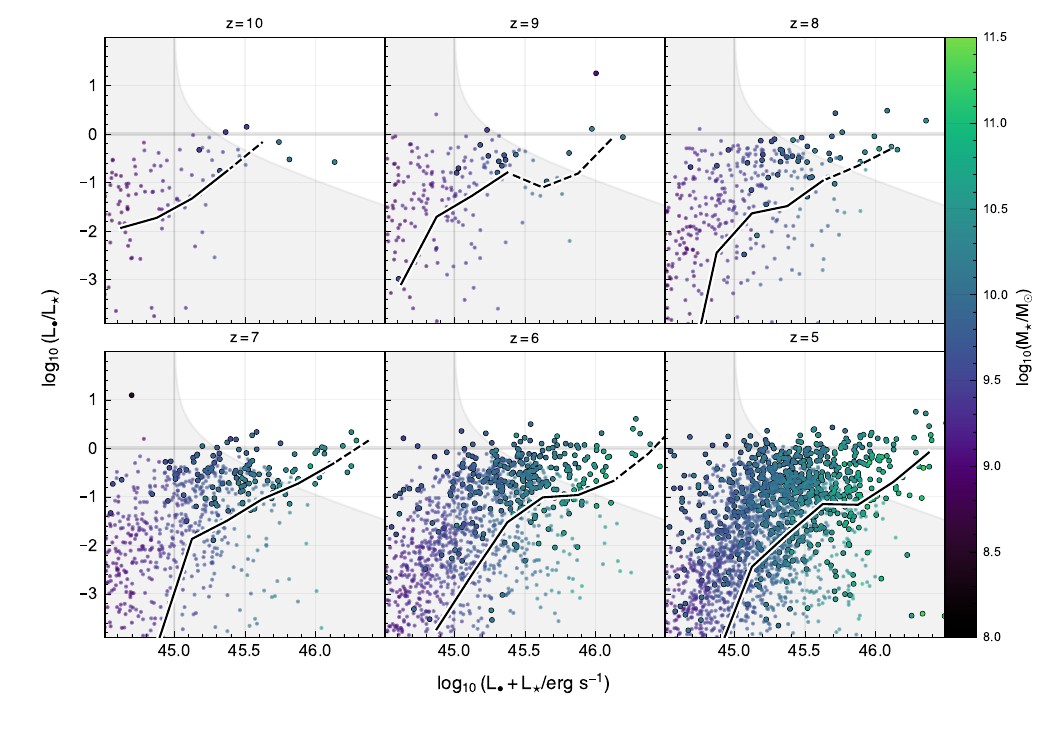}
    \caption{Ratio of SMBH bolometric luminosity to that of the stellar component of \flares\ galaxies at $z\in[5, 10]$ as a function of the total bolometric luminosity. Points are coloured coded by their stellar mass with outlined points those with $M_{\bullet}>10^{7}\ {\rm M_{\odot}}$. The dark solid line denotes the median of this relation with the dashed line denoting where number of objects in the bin are below five. The solid horizontal line gives the 1:1 relation, i.e. where $L_{\star}=L_{\bullet}$. Note, because of the FLARES simulation strategy each galaxy can have a unique weight and thus the median line will not be the median of un-weighted objects. The shaded region denotes where $L_{\bullet}<10^{45}\ {\rm erg\ s^{-1}}$. }
    \label{fig:hosts:bolometric_relative}
\end{figure*}


We next explore the relative contribution of SMBH emission to the total bolometric luminosity of galaxies. In Figure \ref{fig:hosts:bolometric_relative} we present the ratio of the SMBH to stellar bolometric luminosity as a function of the total bolometric luminosity. The bolometric luminosities of objects with ratios $>1$ are then dominated by emission from a SMBH. 


While there is a large amount of scatter in this relation there is a clear trend of an increase in the $L_{\bullet}/L_{\star}$ ratio with total bolometric luminosity.

Objects with $L_{\rm bol}>10^{46.5}\ {\rm erg/s}$ ($>10^{12.9}\ {\rm L_{\odot}}$) on average have SMBH bolometric luminosities surpassing starlight. At least at $5 \le z \le 7$ this transition threshold remains stable. This transition luminosity is similar to the $z=3$ transition luminosity in \citet{Hopkins2010} model but falls short of the transition inferred at $z=4-6$, albeit with a large uncertainty. At $z>7$ the transition may be shifting to lower luminosity, however the number of sources, and therefore the significance, is very low. Below $L_{\rm bol}\sim 10^{45.5}\ {\rm erg/s}$ the average fractional contribution drops significantly. This sharp drops like reflects the choice of halo-seeding mass and seed mass in FLARES. It is also worth highlighting that there is a large amount of scatter at fixed total bolometric luminosity. For example, even at $L_{\rm bol}\approx 10^{45}\ {\rm erg/s}$ at $z=5$, where on average (median) SMBHs only contribute $\approx 0.1\%$ of the total bolometric luminosity, around $1\%$ are dominated by their SMBH.


\subsection{Total Bolometric Luminosity Function}\label{sec:hosts:lf}

\begin{figure*}
    \centering
    \includegraphics[width=2\columnwidth]{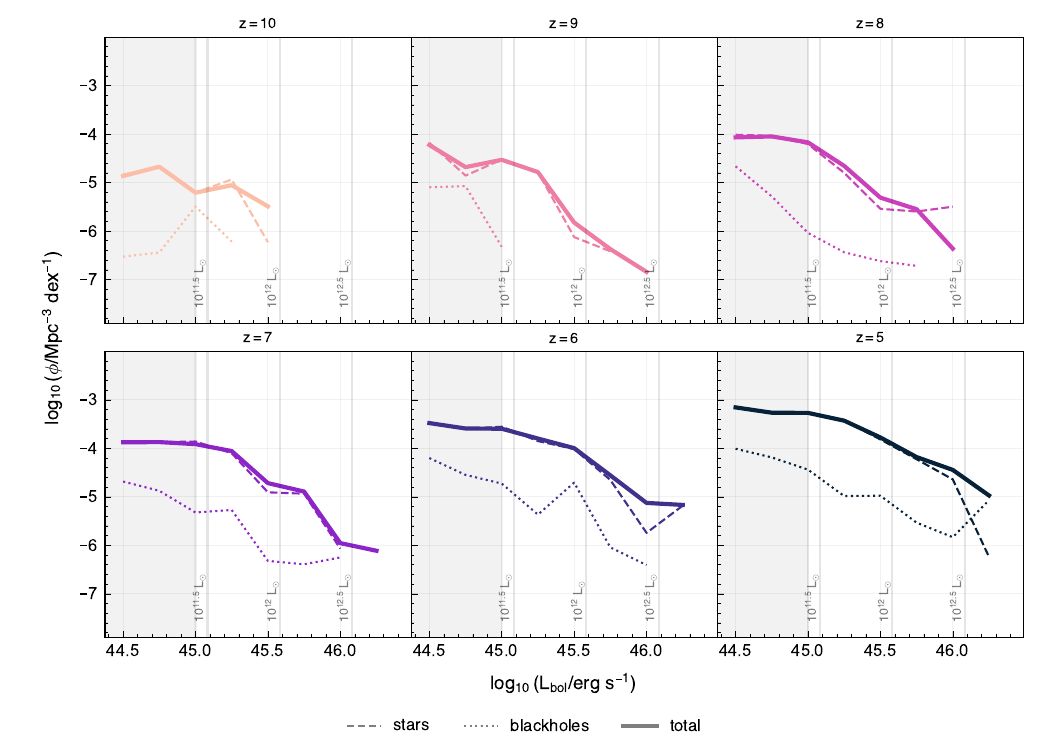}
    \caption{Bolometric luminosity function for stars, SMBHs, and galaxies (combined stars and SMBHs) at $z\in[5, 10]$ predicted by FLARES. Dashed (dotted) lines give the stellar (AGN) contribution to the total galaxy luminosity function given in solid lines.}
    \label{fig:hosts:bol:lf_total}
\end{figure*}

Building on this analysis, in Figure \ref{fig:hosts:bol:lf_total} we show the bolometric function of stars, SMBHs, and the total at $5<z<10$. As anticipated from Figure \ref{fig:hosts:bolometric_relative} the bolometric luminosity function is dominated by stars at $L_{\rm bol}<10^{46}\ {\rm erg/s}$ becoming dominated by SMBHs above, at least at $z=5$.

\subsection{The impact of AGN on their host galaxies}\label{sec:impact}

In EAGLE/FLARES accretion on to SMBHs releases energy, heating neighbouring gas particles. The FLARES simulation strategy makes it easy to experiment with changes to the model. To study the impact of AGN on galaxy formation, we re-simulate one of the high-density regions (region 03, $\delta_{14}\approx -0.31$) but with AGN feedback (see \S\ref{sec:sim:bhs:feedback}) turned off. In this variant SMBHs still grow, but they do not inject energy to the ISM. 

\begin{figure}
    \centering
    \includegraphics[width=\columnwidth]{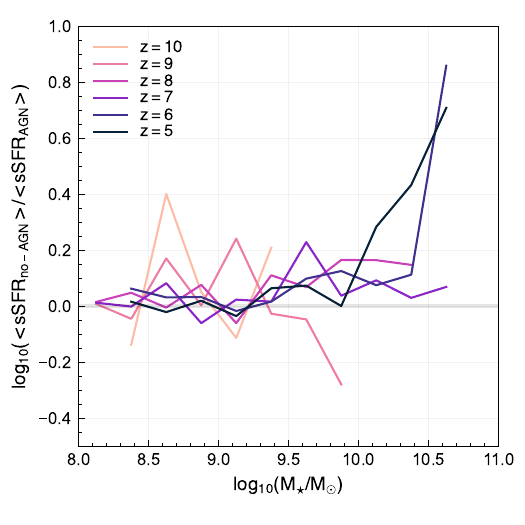}
    \caption{The difference between the mean specific star formation rate with and without AGN feedback activated for galaxies $z=5-10$ for a single simulated FLARES region.}
    \label{fig:impact:ssfr}
\end{figure}

In Figure \ref{fig:impact:ssfr} we show the difference between the mean specific star formation rate in bins of stellar mass. At the highest redshifts this is predictably noisy due to the small number of galaxies in this single simulation. At lower redshift however the number of galaxies has increased enough for us to be confident. This reveals that at low masses ($<10^{9}\ {\rm M_{\odot}}$) there is little or no impact on the \emph{average} specific star formation rates of galaxies. However, there is tentative evidence for a suppression of star formation in the most massive ($>10^{10}\ {\rm M_{\odot}}$) galaxies due to the effect of AGN feedback. 

\citet{FLARES-VIII} studied the emergence of passive galaxies in the early Universe using FLARES.
The EAGLE model produces number densities of passive galaxies in good agreement with observational constraints at $z < 5$ \citep[e.g][]{Merlin2019, Carnall2023}, which gives us confidence in looking at the passive populations at higher redshift.
The main finding in \citet{FLARES-VIII} was that AGN feedback in particular was necessary to produce passive galaxies at $z \geqslant 5$, in agreement with the overall trends in specific star formation rate shown in Figure \ref{fig:impact:ssfr}.
Passive galaxies in FLARES are always those that have the largest SMBHs for their given stellar mass, however, while the growth of SMBHs was found to explain the suppression of star formation in passive galaxies, a large or accreting SMBH doesn't necessarily result in the formation of a passive galaxy. 
Further investigating their SMBH accretion and star formation histories \citet{FLARES-VIII} found that the star formation activity in passive galaxies was anti-correlated with the SMBH accretion rate; passivity tended to follow a period of black hole accretion, and could persist for up to $\sim$400 Myr.

\section{Conclusions}\label{sec:conclusions}

In this work, we have explored the physical and limited photometric properties of super-massive black holes (SMBHs) and their hosts at high-redshift ($5\le z\le 10$) using the First Light And Reionisation Epoch Simulations (FLARES). FLARES is a suite of 40 hydrodynamical zoom simulations employing a variant of the EAGLE physics model. The re-simulations encompass a wide range of  environments ($\delta_{14}\approx -0.3\to 0.3$), with a bias to extreme over-densities, making it ideally suited to studying rare, massive, and luminous objects. As a consequence of this strategy, despite simulating a volume only slightly larger than the EAGLE reference $(100\ {\rm Mpc})^3$ simulation FLARES simulates $\approx 8-20$ times more $M_{\bullet}>10^{7}\ {\rm M_{\odot}}$ SMBHs at $z=5-9$ and $\approx 25$ times more $M_{\bullet}>10^{8}\ {\rm M_{\odot}}$ SMBHs at $z=6$, with samples of  $M_{\bullet}>10^{8}\ {\rm M_{\odot}}$ SMBHs extending to $z=9$. By appropriately weighting each simulation we are able to recover, but also crucially extend, key distribution functions and scaling relations. Our conclusions are:

\begin{itemize}
    
    \item The number density of SMBHs predicted by FLARES drops by $\approx 10\times$ from $z=5\to 7$. This trend may continue to higher-redshift but beyond $z=8$ the number of SMBHs simulated by FLARES is small. The density of SMBHs also drops by $\approx 10\times$ for SMBHs with $M_{\bullet}=10^{7}\to 10^{8}\ {\rm M_{\odot}}$. FLARES predictions are compatible with recent observations of the $z=5$ broad line SMBH mass function; however there is significant uncertainty about the required completeness correction required to convert the broad line mass function to a total SMBH mass function. Where they overlap (in mass and redshift) FLARES is in relatively good agreement with some other models, including Bluetides, Illustris, and TNG100, but lies significantly above models including Astrid, Simba, and TNG300. This reflects the different modelling choices. However, since all the models make multiple changes identifying the decisive factor driving the differences is currently impossible. 

    \item SMBHs are preferentially found in over-dense environments. The densest regions simulated by FLARES ($\delta_{14} = 0.3$) have a density of $M_{\bullet}>10^{7}\ {\rm M_{\odot}}$ SMBHs $\approx 20\times$ higher than mean density regions. 

    \item At fixed mass, SMBHs in FLARES exhibit a range of accretion rates, with almost all having Eddington ratios $10^{-6}\le\lambda_{\rm Edd}\le 1$. The median $\lambda_{\rm Edd}$ is $\approx 10^{-1}$, though decreases with $M_{\bullet}$.

    \item The predicted normalisation of the SMBH bolometric luminosity function evolves by a factor of $\sim 100$ from $z=10\to 5$. At $z=5-6$ it provides a reasonable match to pre-\emph{JWST} constraints. While the $z=5$ luminosity function is consistent with \emph{JWST} observations, at higher redshift there is increasing disagreement, particularly at the bright end with the observational inferred number densities much higher than predicted by FLARES. This tension may have an observational root, for example in the derivation of bolometric luminosities or in the mis-identification of AGN at the highest redshifts. This tension may also reflect modelling choices, for example the choice of radiative efficiency and accretion limit. However, the impact of changing this parameters is complex as enhanced accretion at one epoch may drive a suppression at later times due to increased feedback.

    \item $M_{\bullet}>10^{7}\ {\rm M_{\odot}}$ SMBHs predominantly lie in galaxies with $M_{\star}>10^{9.5}\ {\rm M_{\odot}}$ and have masses $0.05-1.0\%$ of the stellar mass, comparable to local SMBHs. In FLARES no SMBH exceeds more than 2\% of the stellar mass content in a galaxy. For galaxies hosting SMBHs with $L_{\rm bol}>10^{44}\ {\rm erg\ s^{-1}}$ the predicted $M_{\bullet}/M_{\star}$ are well matched to recent \emph{JWST} observations.

    \item The contribution of SMBHs to the bolometric luminosities of galaxies is found to rapidly increase as a function total bolometric luminosity. For galaxies at $5<z<7$ with $L_{\rm bol}>10^{46.5}\, {\rm erg\ s^{-1}}$ we find that accretion on to SMBHs, on average, dominates the total bolometric luminosity.

    \item Using a pair of re-simulations of FLARES regions without AGN feedback enabled, we explore the impact of AGN on their host galaxies. By simply comparing the correlation between the stellar mass and specific star formation rate we find that AGN feedback has the effect of reducing the average star formation activity, but only in the most massive galaxies at the lowest redshifts explored by FLARES ($z=5-6$). In a companion work \citep{FLARES-VIII} we explored the origin of passive galaxies predicted by FLARES finding that their passivity was driven by AGN feedback. 
    
\end{itemize}

\subsection{Future directions}

Looking to the future, \emph{JWST} will continue building up observations - both larger imaging surveys, allowing us to photometrically identify AGN, and spectroscopic follow-up providing the means to unambiguously determine the contribute of AGN in composite objects. Furthermore, the Vera Rubin Observatory and \emph{Euclid} spacecraft are now embarking on their missions to map much of the extragalactic sky across the optical and near-IR. These observations will allow us to identify bright/rare candidate AGN for subsequent follow-up, and crucially confirmation, by \emph{JWST}. Synergies with observatories at other wavelengths will increasingly improve constraints on bolometric luminosities, minimising the need to rely on empirical bolometric corrections obtained at low-redshift.

On the modelling side it is evident that FLARES, through its unique strategy, has provided new insights into early SMBH formation and evolution. However, FLARES is only just probing the regime in which AGN dominate the emission of galaxies. To fully exploit wide-area observational surveys it is essential that even larger simulations are conducted, all while maintaining sufficiently high-resolution. Moreover, except for a handful of tests, FLARES adopted a single physics model and parameter set. Ideally, we would systematically vary the key parameters governing SMBH seeding, growth, dynamics, and feedback to gain clearer insights. We are now preparing for a new phase of the FLARES project in which we will not only employ an updated physics model, but simulate even larger effective volumes and with higher resolution. In the longer term we will also carry out ensemble suites, varying all of the key parameters of the model, and even exploring entirely different modelling approaches.

\newpage

\section*{Changes from Version 1}

In Version 1 of this manuscript we used the number densities presented in the original submission of \citet{Kokorev2024} in Figure \ref{fig:hosts:bol:lf_total}; it was subsequently discovered that, due to an analysis error, these values were erroneously high by $\approx 0.2$ dex. In this version we use updated values from \citet{Kokorev2024} which improves the agreement between our predictions and \citet{Kokorev2024}, though a discrepancy still exists at $z>5$, particularly in the most luminous bins. In the process of addressing comments from the two reviewers we also made several other significant changes. Instead of using the instantaneous accretion rates we decided to use the accretion rates averaged over 10 Myr, shifting the discussion of this from the Appendix to the main body of the text. This has a subtle impact on all Figures from Figure \ref{fig:bolometric_luminosity_function-averaging} on-wards. We also uncovered a bug in the calculation of stellar bolometric luminosities which led them to be under-reported. This has a significant impact on what are now Figures \ref{fig:hosts:bolometric_relative} and \ref{fig:hosts:bol:lf_total}. 

\section*{Author Contributions}

We list here the roles and contributions of the authors according to the Contributor Roles Taxonomy (CRediT)\footnote{\url{https://credit.niso.org/}}.
\textbf{Jussi K. Kuusisto, Stephen M. Wilkins}: Conceptualization, Data curation, Methodology, Investigation, Formal Analysis, Visualization, Writing - original draft.
\textbf{Christopher C. Lovell}: Conceptualization, Data curation, Methodology, Writing - original draft.
\textbf{Dimitrios Irodotou, Shihong Liao, Sonja Soininen}: Investigation, Writing - original draft.
\textbf{William Roper, Aswin P. Vijayan}: Data curation, review \& editing.
\textbf{Peter A. Thomas} Conceptualization, Writing - review \& editing.
\textbf{Sabrina C. Berger, Sophie L. Newman, Louise T. C. Seeyave, Shihong Liao}: Writing - review \& editing.

\section*{Acknowledgements}

We would like to thank the two referees for their comprehensive reviews which we believe have resulted in an significantly improved manuscript. We thank the \eagle\, team for their efforts in developing the \eagle\, simulation code.  This work used the DiRAC@Durham facility managed by the Institute for Computational Cosmology on behalf of the STFC DiRAC HPC Facility (www.dirac.ac.uk). The equipment was funded by BEIS capital funding via STFC capital grants ST/K00042X/1, ST/P002293/1, ST/R002371/1 and ST/S002502/1, Durham University and STFC operations grant ST/R000832/1. DiRAC is part of the National e-Infrastructure. 

JKK and LTCS are supported by an STFC studentship. SMW and WJR acknowledge support from the Sussex Astronomy Centre STFC Consolidated Grant (ST/X001040/1). CCL acknowledges support from a Dennis Sciama fellowship funded by the University of Portsmouth for the Institute of Cosmology and Gravitation. DI acknowledges support by the European Research Council via ERC Consolidator Grant KETJU (no. 818930) and the CSC – IT Center for Science, Finland. APV acknowledges support from the Carlsberg Foundation (grant no CF20-0534). The Cosmic Dawn Center (DAWN) is funded by the Danish National Research Foundation under grant No. 140. SL acknowledges the supports by the National Natural Science Foundation of China (NSFC) grant (No. 11988101) and the K. C. Wong Education Foundation.

 We also wish to acknowledge the following open source software packages used in the analysis: \textsc{Numpy} \citep{numpy}, \textsc{Scipy} \citep{scipy}, \textsc{Astropy} \citep{astropy:2013, astropy:2018, astropy:2022}, \textsc{Cmasher} \citep{cmasher}, and \textsc{Matplotlib} \citep{matplotlib}. 

\section*{Data Availability Statement}

The data associated with the paper will be made publicly available at \href{https://flaresimulations.github.io}{https://flaresimulations.github.io} on the acceptance of the manuscript.



\bibliographystyle{mnras}
\bibliography{flares_agn, flares} 

\begin{thebibliography}{}
\makeatletter
\relax
\def\mn@urlcharsother{\let\do\@makeother \do\$\do\&\do\#\do\^\do\_\do\%\do\~}
\def\mn@doi{\begingroup\mn@urlcharsother \@ifnextchar [ {\mn@doi@} {\mn@doi@[]}}
\def\mn@doi@[#1]#2{\def\@tempa{#1}\ifx\@tempa\@empty \href {http://dx.doi.org/#2} {doi:#2}\else \href {http://dx.doi.org/#2} {#1}\fi \endgroup}
\def\mn@eprint#1#2{\mn@eprint@#1:#2::\@nil}
\def\mn@eprint@arXiv#1{\href {http://arxiv.org/abs/#1} {{\tt arXiv:#1}}}
\def\mn@eprint@dblp#1{\href {http://dblp.uni-trier.de/rec/bibtex/#1.xml} {dblp:#1}}
\def\mn@eprint@#1:#2:#3:#4\@nil{\def\@tempa {#1}\def\@tempb {#2}\def\@tempc {#3}\ifx \@tempc \@empty \let \@tempc \@tempb \let \@tempb \@tempa \fi \ifx \@tempb \@empty \def\@tempb {arXiv}\fi \@ifundefined {mn@eprint@\@tempb}{\@tempb:\@tempc}{\expandafter \expandafter \csname mn@eprint@\@tempb\endcsname \expandafter{\@tempc}}}

\bibitem[\protect\citeauthoryear{{Abramowicz}, {Czerny}, {Lasota}  \& {Szuszkiewicz}}{{Abramowicz} et~al.}{1988}]{Abramowicz1988}
{Abramowicz} M.~A.,  {Czerny} B.,  {Lasota} J.~P.,   {Szuszkiewicz} E.,  1988, \mn@doi [\apj] {10.1086/166683}, \href {https://ui.adsabs.harvard.edu/abs/1988ApJ...332..646A} {332, 646}

\bibitem[\protect\citeauthoryear{{Astropy Collaboration} et~al.,}{{Astropy Collaboration} et~al.}{2013}]{astropy:2013}
{Astropy Collaboration} et~al., 2013, \mn@doi [\aap] {10.1051/0004-6361/201322068}, \href {http://adsabs.harvard.edu/abs/2013A%26A...558A..33A} {558, A33}

\bibitem[\protect\citeauthoryear{{Astropy Collaboration} et~al.,}{{Astropy Collaboration} et~al.}{2018}]{astropy:2018}
{Astropy Collaboration} et~al., 2018, \mn@doi [\aj] {10.3847/1538-3881/aabc4f}, \href {https://ui.adsabs.harvard.edu/abs/2018AJ....156..123A} {156, 123}

\bibitem[\protect\citeauthoryear{{Astropy Collaboration} et~al.,}{{Astropy Collaboration} et~al.}{2022}]{astropy:2022}
{Astropy Collaboration} et~al., 2022, \mn@doi [\apj] {10.3847/1538-4357/ac7c74}, \href {https://ui.adsabs.harvard.edu/abs/2022ApJ...935..167A} {935, 167}

\bibitem[\protect\citeauthoryear{{Bah{\'e}} et~al.,}{{Bah{\'e}} et~al.}{2022}]{Bahe2022}
{Bah{\'e}} Y.~M.,  et~al., 2022, \mn@doi [\mnras] {10.1093/mnras/stac1339}, \href {https://ui.adsabs.harvard.edu/abs/2022MNRAS.516..167B} {516, 167}

\bibitem[\protect\citeauthoryear{{Bardeen}, {Carter}  \& {Hawking}}{{Bardeen} et~al.}{1973}]{Bardeen1973}
{Bardeen} J.~M.,  {Carter} B.,   {Hawking} S.~W.,  1973, \mn@doi [Communications in Mathematical Physics] {10.1007/BF01645742}, \href {https://ui.adsabs.harvard.edu/abs/1973CMaPh..31..161B} {31, 161}

\bibitem[\protect\citeauthoryear{{Barnes} et~al.,}{{Barnes} et~al.}{2017}]{CEAGLE}
{Barnes} D.~J.,  et~al., 2017, \mn@doi [\mnras] {10.1093/mnras/stx1647}, \href {https://ui.adsabs.harvard.edu/abs/2017MNRAS.471.1088B} {471, 1088}

\bibitem[\protect\citeauthoryear{{Begelman}, {Volonteri}  \& {Rees}}{{Begelman} et~al.}{2006}]{Begelman2006}
{Begelman} M.~C.,  {Volonteri} M.,   {Rees} M.~J.,  2006, \mn@doi [\mnras] {10.1111/j.1365-2966.2006.10467.x}, \href {https://ui.adsabs.harvard.edu/abs/2006MNRAS.370..289B} {370, 289}

\bibitem[\protect\citeauthoryear{{Bird}, {Ni}, {Di Matteo}, {Croft}, {Feng}  \& {Chen}}{{Bird} et~al.}{2022}]{Bird2022}
{Bird} S.,  {Ni} Y.,  {Di Matteo} T.,  {Croft} R.,  {Feng} Y.,   {Chen} N.,  2022, \mn@doi [\mnras] {10.1093/mnras/stac648}, \href {https://ui.adsabs.harvard.edu/abs/2022MNRAS.512.3703B} {512, 3703}

\bibitem[\protect\citeauthoryear{{Blandford} \& {Znajek}}{{Blandford} \& {Znajek}}{1977}]{Blandford1977}
{Blandford} R.~D.,  {Znajek} R.~L.,  1977, \mn@doi [\mnras] {10.1093/mnras/179.3.433}, \href {https://ui.adsabs.harvard.edu/abs/1977MNRAS.179..433B} {179, 433}

\bibitem[\protect\citeauthoryear{{Bogd{\'a}n} et~al.,}{{Bogd{\'a}n} et~al.}{2023}]{Bogdan2023}
{Bogd{\'a}n} {\'A}.,  et~al., 2023, \mn@doi [Nature Astronomy] {10.1038/s41550-023-02111-9}, \href {https://ui.adsabs.harvard.edu/abs/2023NatAs.tmp..223B} {}

\bibitem[\protect\citeauthoryear{{Bondi} \& {Hoyle}}{{Bondi} \& {Hoyle}}{1944}]{Bondi1944}
{Bondi} H.,  {Hoyle} F.,  1944, \mn@doi [\mnras] {10.1093/mnras/104.5.273}, \href {https://ui.adsabs.harvard.edu/abs/1944MNRAS.104..273B} {104, 273}

\bibitem[\protect\citeauthoryear{{Booth} \& {Schaye}}{{Booth} \& {Schaye}}{2009}]{Booth2009}
{Booth} C.~M.,  {Schaye} J.,  2009, \mn@doi [\mnras] {10.1111/j.1365-2966.2009.15043.x}, \href {https://ui.adsabs.harvard.edu/abs/2009MNRAS.398...53B} {398, 53}

\bibitem[\protect\citeauthoryear{{Bower}, {Benson}, {Malbon}, {Helly}, {Frenk}, {Baugh}, {Cole}  \& {Lacey}}{{Bower} et~al.}{2006}]{Bower2006}
{Bower} R.~G.,  {Benson} A.~J.,  {Malbon} R.,  {Helly} J.~C.,  {Frenk} C.~S.,  {Baugh} C.~M.,  {Cole} S.,   {Lacey} C.~G.,  2006, \mn@doi [\mnras] {10.1111/j.1365-2966.2006.10519.x}, \href {https://ui.adsabs.harvard.edu/abs/2006MNRAS.370..645B} {370, 645}

\bibitem[\protect\citeauthoryear{{Bromm} \& {Loeb}}{{Bromm} \& {Loeb}}{2003}]{Bromm2003}
{Bromm} V.,  {Loeb} A.,  2003, \mn@doi [\apj] {10.1086/377529}, \href {https://ui.adsabs.harvard.edu/abs/2003ApJ...596...34B} {596, 34}

\bibitem[\protect\citeauthoryear{{Carnall} et~al.,}{{Carnall} et~al.}{2023}]{Carnall2023}
{Carnall} A.~C.,  et~al., 2023, \mn@doi [\mnras] {10.1093/mnras/stad369}, \href {https://ui.adsabs.harvard.edu/abs/2023MNRAS.520.3974C} {520, 3974}

\bibitem[\protect\citeauthoryear{{Cattaneo}, {Dekel}, {Devriendt}, {Guiderdoni}  \& {Blaizot}}{{Cattaneo} et~al.}{2006}]{Cattaneo2006}
{Cattaneo} A.,  {Dekel} A.,  {Devriendt} J.,  {Guiderdoni} B.,   {Blaizot} J.,  2006, \mn@doi [\mnras] {10.1111/j.1365-2966.2006.10608.x}, \href {https://ui.adsabs.harvard.edu/abs/2006MNRAS.370.1651C} {370, 1651}

\bibitem[\protect\citeauthoryear{Chabrier}{Chabrier}{2003}]{ChabrierIMF}
Chabrier G.,  2003, \mn@doi [\pasp] {10.1086/376392}, 115, 763

\bibitem[\protect\citeauthoryear{{Choi}, {Ostriker}, {Naab}, {Oser}  \& {Moster}}{{Choi} et~al.}{2015}]{Choi2015}
{Choi} E.,  {Ostriker} J.~P.,  {Naab} T.,  {Oser} L.,   {Moster} B.~P.,  2015, \mn@doi [\mnras] {10.1093/mnras/stv575}, \href {https://ui.adsabs.harvard.edu/abs/2015MNRAS.449.4105C} {449, 4105}

\bibitem[\protect\citeauthoryear{{Ciotti} \& {Ostriker}}{{Ciotti} \& {Ostriker}}{2001}]{Ciotti2001}
{Ciotti} L.,  {Ostriker} J.~P.,  2001, \mn@doi [\apj] {10.1086/320053}, \href {https://ui.adsabs.harvard.edu/abs/2001ApJ...551..131C} {551, 131}

\bibitem[\protect\citeauthoryear{{Costa}, {Sijacki}  \& {Haehnelt}}{{Costa} et~al.}{2014}]{Costa2014}
{Costa} T.,  {Sijacki} D.,   {Haehnelt} M.~G.,  2014, \mn@doi [\mnras] {10.1093/mnras/stu1632}, \href {https://ui.adsabs.harvard.edu/abs/2014MNRAS.444.2355C} {444, 2355}

\bibitem[\protect\citeauthoryear{{Costa}, {Pakmor}  \& {Springel}}{{Costa} et~al.}{2020}]{Costa2020}
{Costa} T.,  {Pakmor} R.,   {Springel} V.,  2020, \mn@doi [\mnras] {10.1093/mnras/staa2321}, \href {https://ui.adsabs.harvard.edu/abs/2020MNRAS.497.5229C} {497, 5229}

\bibitem[\protect\citeauthoryear{Crain et~al.,}{Crain et~al.}{2015}]{crain_eagle_2015}
Crain R.~A.,  et~al., 2015, \mn@doi [\mnras] {10.1093/mnras/stv725}, 450, 1937

\bibitem[\protect\citeauthoryear{{Croton} et~al.,}{{Croton} et~al.}{2006}]{Croton2006}
{Croton} D.~J.,  et~al., 2006, \mn@doi [\mnras] {10.1111/j.1365-2966.2005.09675.x}, \href {https://ui.adsabs.harvard.edu/abs/2006MNRAS.365...11C} {365, 11}

\bibitem[\protect\citeauthoryear{{Dalla Vecchia} \& {Schaye}}{{Dalla Vecchia} \& {Schaye}}{2008}]{DallaVecchia2008}
{Dalla Vecchia} C.,  {Schaye} J.,  2008, \mn@doi [\mnras] {10.1111/j.1365-2966.2008.13322.x}, \href {https://ui.adsabs.harvard.edu/abs/2008MNRAS.387.1431D} {387, 1431}

\bibitem[\protect\citeauthoryear{{Dav{\'e}}, {Angl{\'e}s-Alc{\'a}zar}, {Narayanan}, {Li}, {Rafieferantsoa}  \& {Appleby}}{{Dav{\'e}} et~al.}{2019}]{Dave2019}
{Dav{\'e}} R.,  {Angl{\'e}s-Alc{\'a}zar} D.,  {Narayanan} D.,  {Li} Q.,  {Rafieferantsoa} M.~H.,   {Appleby} S.,  2019, \mn@doi [\mnras] {10.1093/mnras/stz937}, \href {https://ui.adsabs.harvard.edu/abs/2019MNRAS.486.2827D} {486, 2827}

\bibitem[\protect\citeauthoryear{{Dayal} \& {Ferrara}}{{Dayal} \& {Ferrara}}{2018}]{Dayal2018}
{Dayal} P.,  {Ferrara} A.,  2018, \mn@doi [\physrep] {10.1016/j.physrep.2018.10.002}, \href {https://ui.adsabs.harvard.edu/abs/2018PhR...780....1D} {780, 1}

\bibitem[\protect\citeauthoryear{{Di Matteo}, {Springel}  \& {Hernquist}}{{Di Matteo} et~al.}{2005}]{DiMatteo2005}
{Di Matteo} T.,  {Springel} V.,   {Hernquist} L.,  2005, \mn@doi [\nat] {10.1038/nature03335}, \href {https://ui.adsabs.harvard.edu/abs/2005Natur.433..604D} {433, 604}

\bibitem[\protect\citeauthoryear{{Di Matteo}, {Croft}, {Feng}, {Waters}  \& {Wilkins}}{{Di Matteo} et~al.}{2017}]{DiMatteo2017}
{Di Matteo} T.,  {Croft} R. A.~C.,  {Feng} Y.,  {Waters} D.,   {Wilkins} S.,  2017, \mn@doi [\mnras] {10.1093/mnras/stx319}, \href {https://ui.adsabs.harvard.edu/abs/2017MNRAS.467.4243D} {467, 4243}

\bibitem[\protect\citeauthoryear{{Dubois}, {Devriendt}, {Slyz}  \& {Teyssier}}{{Dubois} et~al.}{2012}]{Dubois2012}
{Dubois} Y.,  {Devriendt} J.,  {Slyz} A.,   {Teyssier} R.,  2012, \mn@doi [\mnras] {10.1111/j.1365-2966.2011.20236.x}, \href {https://ui.adsabs.harvard.edu/abs/2012MNRAS.420.2662D} {420, 2662}

\bibitem[\protect\citeauthoryear{{Dubois}, {Peirani}, {Pichon}, {Devriendt}, {Gavazzi}, {Welker}  \& {Volonteri}}{{Dubois} et~al.}{2016}]{Dubois2016}
{Dubois} Y.,  {Peirani} S.,  {Pichon} C.,  {Devriendt} J.,  {Gavazzi} R.,  {Welker} C.,   {Volonteri} M.,  2016, \mn@doi [\mnras] {10.1093/mnras/stw2265}, \href {https://ui.adsabs.harvard.edu/abs/2016MNRAS.463.3948D} {463, 3948}

\bibitem[\protect\citeauthoryear{{Eilers} et~al.,}{{Eilers} et~al.}{2024}]{Eilers2024}
{Eilers} A.-C.,  et~al., 2024, \mn@doi [\apj] {10.3847/1538-4357/ad778b}, \href {https://ui.adsabs.harvard.edu/abs/2024ApJ...974..275E} {974, 275}

\bibitem[\protect\citeauthoryear{{Fabian}}{{Fabian}}{2012}]{Fabian2012}
{Fabian} A.~C.,  2012, \mn@doi [\araa] {10.1146/annurev-astro-081811-125521}, \href {https://ui.adsabs.harvard.edu/abs/2012ARA&A..50..455F} {50, 455}

\bibitem[\protect\citeauthoryear{{Fan}, {Ba{\~n}ados}  \& {Simcoe}}{{Fan} et~al.}{2023}]{Fan2023}
{Fan} X.,  {Ba{\~n}ados} E.,   {Simcoe} R.~A.,  2023, \mn@doi [\araa] {10.1146/annurev-astro-052920-102455}, \href {https://ui.adsabs.harvard.edu/abs/2023ARA&A..61..373F} {61, 373}

\bibitem[\protect\citeauthoryear{{Ferrarese} \& {Merritt}}{{Ferrarese} \& {Merritt}}{2000}]{Ferrarese2000}
{Ferrarese} L.,  {Merritt} D.,  2000, \mn@doi [\apjl] {10.1086/312838}, \href {https://ui.adsabs.harvard.edu/abs/2000ApJ...539L...9F} {539, L9}

\bibitem[\protect\citeauthoryear{{Finkelstein} \& {Bagley}}{{Finkelstein} \& {Bagley}}{2022}]{Finkelstein2022}
{Finkelstein} S.~L.,  {Bagley} M.~B.,  2022, \mn@doi [\apj] {10.3847/1538-4357/ac89eb}, \href {https://ui.adsabs.harvard.edu/abs/2022ApJ...938...25F} {938, 25}

\bibitem[\protect\citeauthoryear{{Gebhardt} et~al.,}{{Gebhardt} et~al.}{2000}]{Gebhardt2000}
{Gebhardt} K.,  et~al., 2000, \mn@doi [\apjl] {10.1086/312840}, \href {https://ui.adsabs.harvard.edu/abs/2000ApJ...539L..13G} {539, L13}

\bibitem[\protect\citeauthoryear{{Genel} et~al.,}{{Genel} et~al.}{2014}]{Genel2014}
{Genel} S.,  et~al., 2014, \mn@doi [\mnras] {10.1093/mnras/stu1654}, \href {https://ui.adsabs.harvard.edu/abs/2014MNRAS.445..175G} {445, 175}

\bibitem[\protect\citeauthoryear{{Giallongo} et~al.,}{{Giallongo} et~al.}{2019}]{Giallongo2019}
{Giallongo} E.,  et~al., 2019, \mn@doi [\apj] {10.3847/1538-4357/ab39e1}, \href {https://ui.adsabs.harvard.edu/abs/2019ApJ...884...19G} {884, 19}

\bibitem[\protect\citeauthoryear{{Granato}, {De Zotti}, {Silva}, {Bressan}  \& {Danese}}{{Granato} et~al.}{2004}]{Granato2004}
{Granato} G.~L.,  {De Zotti} G.,  {Silva} L.,  {Bressan} A.,   {Danese} L.,  2004, \mn@doi [\apj] {10.1086/379875}, \href {https://ui.adsabs.harvard.edu/abs/2004ApJ...600..580G} {600, 580}

\bibitem[\protect\citeauthoryear{{Greene} et~al.,}{{Greene} et~al.}{2023}]{Greene23}
{Greene} J.~E.,  et~al., 2023, \mn@doi [arXiv e-prints] {10.48550/arXiv.2309.05714}, \href {https://ui.adsabs.harvard.edu/abs/2023arXiv230905714G} {p. arXiv:2309.05714}

\bibitem[\protect\citeauthoryear{{Haardt} \& {Madau}}{{Haardt} \& {Madau}}{2001}]{Haardt2001}
{Haardt} F.,  {Madau} P.,  2001, in {Neumann} D.~M.,  {Tran} J.~T.~V.,  eds, Clusters of Galaxies and the High Redshift Universe Observed in X-rays. p.~64 (\mn@eprint {arXiv} {astro-ph/0106018}), \mn@doi{10.48550/arXiv.astro-ph/0106018}

\bibitem[\protect\citeauthoryear{{Habouzit} et~al.,}{{Habouzit} et~al.}{2022a}]{Habouzit2022a}
{Habouzit} M.,  et~al., 2022a, \mn@doi [\mnras] {10.1093/mnras/stab3147}, \href {https://ui.adsabs.harvard.edu/abs/2022MNRAS.509.3015H} {509, 3015}

\bibitem[\protect\citeauthoryear{{Habouzit} et~al.,}{{Habouzit} et~al.}{2022b}]{Habouzit2022b}
{Habouzit} M.,  et~al., 2022b, \mn@doi [\mnras] {10.1093/mnras/stac225}, \href {https://ui.adsabs.harvard.edu/abs/2022MNRAS.511.3751H} {511, 3751}

\bibitem[\protect\citeauthoryear{{Harikane} et~al.,}{{Harikane} et~al.}{2023}]{Harikane2023}
{Harikane} Y.,  et~al., 2023, \mn@doi [\apj] {10.3847/1538-4357/ad029e}, \href {https://ui.adsabs.harvard.edu/abs/2023ApJ...959...39H} {959, 39}

\bibitem[\protect\citeauthoryear{{H{\"a}ring} \& {Rix}}{{H{\"a}ring} \& {Rix}}{2004}]{Haring2004}
{H{\"a}ring} N.,  {Rix} H.-W.,  2004, \mn@doi [\apjl] {10.1086/383567}, \href {https://ui.adsabs.harvard.edu/abs/2004ApJ...604L..89H} {604, L89}

\bibitem[\protect\citeauthoryear{Harris et~al.,}{Harris et~al.}{2020}]{numpy}
Harris C.~R.,  et~al., 2020, \mn@doi [Nature] {10.1038/s41586-020-2649-2}, 585, 357

\bibitem[\protect\citeauthoryear{{He} et~al.,}{{He} et~al.}{2023}]{He2023}
{He} W.,  et~al., 2023, \mn@doi [arXiv e-prints] {10.48550/arXiv.2311.08922}, \href {https://ui.adsabs.harvard.edu/abs/2023arXiv231108922H} {p. arXiv:2311.08922}

\bibitem[\protect\citeauthoryear{{Hopkins}, {Hernquist}, {Cox}, {Di Matteo}, {Robertson}  \& {Springel}}{{Hopkins} et~al.}{2006}]{Hopkins2006}
{Hopkins} P.~F.,  {Hernquist} L.,  {Cox} T.~J.,  {Di Matteo} T.,  {Robertson} B.,   {Springel} V.,  2006, \mn@doi [\apjs] {10.1086/499298}, \href {https://ui.adsabs.harvard.edu/abs/2006ApJS..163....1H} {163, 1}

\bibitem[\protect\citeauthoryear{{Hopkins}, {Younger}, {Hayward}, {Narayanan}  \& {Hernquist}}{{Hopkins} et~al.}{2010}]{Hopkins2010}
{Hopkins} P.~F.,  {Younger} J.~D.,  {Hayward} C.~C.,  {Narayanan} D.,   {Hernquist} L.,  2010, \mn@doi [\mnras] {10.1111/j.1365-2966.2009.15990.x}, \href {https://ui.adsabs.harvard.edu/abs/2010MNRAS.402.1693H} {402, 1693}

\bibitem[\protect\citeauthoryear{{Huang}, {Di Matteo}, {Bhowmick}, {Feng}  \& {Ma}}{{Huang} et~al.}{2018}]{Huang2018}
{Huang} K.-W.,  {Di Matteo} T.,  {Bhowmick} A.~K.,  {Feng} Y.,   {Ma} C.-P.,  2018, \mn@doi [\mnras] {10.1093/mnras/sty1329}, \href {https://ui.adsabs.harvard.edu/abs/2018MNRAS.478.5063H} {478, 5063}

\bibitem[\protect\citeauthoryear{Hunter}{Hunter}{2007}]{matplotlib}
Hunter J.~D.,  2007, \mn@doi [Computing in Science \& Engineering] {10.1109/MCSE.2007.55}, 9, 90

\bibitem[\protect\citeauthoryear{{Inayoshi}, {Visbal}  \& {Haiman}}{{Inayoshi} et~al.}{2020}]{Inayoshi2020}
{Inayoshi} K.,  {Visbal} E.,   {Haiman} Z.,  2020, \mn@doi [\araa] {10.1146/annurev-astro-120419-014455}, \href {https://ui.adsabs.harvard.edu/abs/2020ARA&A..58...27I} {58, 27}

\bibitem[\protect\citeauthoryear{{Jiang} et~al.,}{{Jiang} et~al.}{2016}]{Jiang2016}
{Jiang} L.,  et~al., 2016, \mn@doi [\apj] {10.3847/1538-4357/833/2/222}, \href {https://ui.adsabs.harvard.edu/abs/2016ApJ...833..222J} {833, 222}

\bibitem[\protect\citeauthoryear{{Juod{\v{z}}balis} et~al.,}{{Juod{\v{z}}balis} et~al.}{2023}]{Juodzbalis2023}
{Juod{\v{z}}balis} I.,  et~al., 2023, \mn@doi [\mnras] {10.1093/mnras/stad2396}, \href {https://ui.adsabs.harvard.edu/abs/2023MNRAS.525.1353J} {525, 1353}

\bibitem[\protect\citeauthoryear{{Kaspi}, {Smith}, {Netzer}, {Maoz}, {Jannuzi}  \& {Giveon}}{{Kaspi} et~al.}{2000}]{Kaspi2000}
{Kaspi} S.,  {Smith} P.~S.,  {Netzer} H.,  {Maoz} D.,  {Jannuzi} B.~T.,   {Giveon} U.,  2000, \mn@doi [\apj] {10.1086/308704}, \href {https://ui.adsabs.harvard.edu/abs/2000ApJ...533..631K} {533, 631}

\bibitem[\protect\citeauthoryear{{Kauffmann} \& {Haehnelt}}{{Kauffmann} \& {Haehnelt}}{2000}]{Kauffmann2000}
{Kauffmann} G.,  {Haehnelt} M.,  2000, \mn@doi [\mnras] {10.1046/j.1365-8711.2000.03077.x}, \href {https://ui.adsabs.harvard.edu/abs/2000MNRAS.311..576K} {311, 576}

\bibitem[\protect\citeauthoryear{{Kerr}}{{Kerr}}{1963}]{Kerr1963}
{Kerr} R.~P.,  1963, \mn@doi [\prl] {10.1103/PhysRevLett.11.237}, \href {https://ui.adsabs.harvard.edu/abs/1963PhRvL..11..237K} {11, 237}

\bibitem[\protect\citeauthoryear{{Khandai}, {Feng}, {DeGraf}, {Di Matteo}  \& {Croft}}{{Khandai} et~al.}{2012}]{Khandai2012}
{Khandai} N.,  {Feng} Y.,  {DeGraf} C.,  {Di Matteo} T.,   {Croft} R. A.~C.,  2012, \mn@doi [\mnras] {10.1111/j.1365-2966.2012.21047.x}, \href {https://ui.adsabs.harvard.edu/abs/2012MNRAS.423.2397K} {423, 2397}

\bibitem[\protect\citeauthoryear{{Kocevski} et~al.,}{{Kocevski} et~al.}{2023}]{Kocevski23}
{Kocevski} D.~D.,  et~al., 2023, \mn@doi [\apjl] {10.3847/2041-8213/ace5a0}, \href {https://ui.adsabs.harvard.edu/abs/2023ApJ...954L...4K} {954, L4}

\bibitem[\protect\citeauthoryear{{Kokorev} et~al.,}{{Kokorev} et~al.}{2024}]{Kokorev2024}
{Kokorev} V.,  et~al., 2024, \mn@doi [arXiv e-prints] {10.48550/arXiv.2401.09981}, \href {https://ui.adsabs.harvard.edu/abs/2024arXiv240109981K} {p. arXiv:2401.09981}

\bibitem[\protect\citeauthoryear{{Kormendy} \& {Ho}}{{Kormendy} \& {Ho}}{2013}]{Kormendy2013}
{Kormendy} J.,  {Ho} L.~C.,  2013, \mn@doi [\araa] {10.1146/annurev-astro-082708-101811}, \href {https://ui.adsabs.harvard.edu/abs/2013ARA&A..51..511K} {51, 511}

\bibitem[\protect\citeauthoryear{{Kormendy} \& {Richstone}}{{Kormendy} \& {Richstone}}{1995}]{Kormendy1995}
{Kormendy} J.,  {Richstone} D.,  1995, \mn@doi [\araa] {10.1146/annurev.aa.33.090195.003053}, \href {https://ui.adsabs.harvard.edu/abs/1995ARA&A..33..581K} {33, 581}

\bibitem[\protect\citeauthoryear{{Kubota} \& {Done}}{{Kubota} \& {Done}}{2018}]{KD18}
{Kubota} A.,  {Done} C.,  2018, \mn@doi [\mnras] {10.1093/mnras/sty1890}, \href {https://ui.adsabs.harvard.edu/abs/2018MNRAS.480.1247K} {480, 1247}

\bibitem[\protect\citeauthoryear{{Larson} et~al.,}{{Larson} et~al.}{2023}]{Larson2023}
{Larson} R.~L.,  et~al., 2023, \mn@doi [\apjl] {10.3847/2041-8213/ace619}, \href {https://ui.adsabs.harvard.edu/abs/2023ApJ...953L..29L} {953, L29}

\bibitem[\protect\citeauthoryear{{Latif}, {Schleicher}, {Schmidt}  \& {Niemeyer}}{{Latif} et~al.}{2013}]{Latif2013}
{Latif} M.~A.,  {Schleicher} D.~R.~G.,  {Schmidt} W.,   {Niemeyer} J.,  2013, \mn@doi [\mnras] {10.1093/mnras/stt834}, \href {https://ui.adsabs.harvard.edu/abs/2013MNRAS.433.1607L} {433, 1607}

\bibitem[\protect\citeauthoryear{{Loeb} \& {Rasio}}{{Loeb} \& {Rasio}}{1994}]{Loeb1994}
{Loeb} A.,  {Rasio} F.~A.,  1994, \mn@doi [\apj] {10.1086/174548}, \href {https://ui.adsabs.harvard.edu/abs/1994ApJ...432...52L} {432, 52}

\bibitem[\protect\citeauthoryear{{Lovell}, {Vijayan}, {Thomas}, {Wilkins}, {Barnes}, {Irodotou}  \& {Roper}}{{Lovell} et~al.}{2021}]{FLARES-I}
{Lovell} C.~C.,  {Vijayan} A.~P.,  {Thomas} P.~A.,  {Wilkins} S.~M.,  {Barnes} D.~J.,  {Irodotou} D.,   {Roper} W.,  2021, \mn@doi [\mnras] {10.1093/mnras/staa3360}, \href {https://ui.adsabs.harvard.edu/abs/2021MNRAS.500.2127L} {500, 2127}

\bibitem[\protect\citeauthoryear{{Lovell} et~al.,}{{Lovell} et~al.}{2023}]{FLARES-VIII}
{Lovell} C.~C.,  et~al., 2023, \mn@doi [\mnras] {10.1093/mnras/stad2550}, \href {https://ui.adsabs.harvard.edu/abs/2023MNRAS.525.5520L} {525, 5520}

\bibitem[\protect\citeauthoryear{{Lynden-Bell}}{{Lynden-Bell}}{1969}]{LyndenBell1969}
{Lynden-Bell} D.,  1969, \mn@doi [\nat] {10.1038/223690a0}, \href {https://ui.adsabs.harvard.edu/abs/1969Natur.223..690L} {223, 690}

\bibitem[\protect\citeauthoryear{{Madau} \& {Haardt}}{{Madau} \& {Haardt}}{2015}]{Madau2015}
{Madau} P.,  {Haardt} F.,  2015, \mn@doi [\apjl] {10.1088/2041-8205/813/1/L8}, \href {https://ui.adsabs.harvard.edu/abs/2015ApJ...813L...8M} {813, L8}

\bibitem[\protect\citeauthoryear{{Madau} \& {Rees}}{{Madau} \& {Rees}}{2001}]{Madau2001}
{Madau} P.,  {Rees} M.~J.,  2001, \mn@doi [\apjl] {10.1086/319848}, \href {https://ui.adsabs.harvard.edu/abs/2001ApJ...551L..27M} {551, L27}

\bibitem[\protect\citeauthoryear{{Magorrian} et~al.,}{{Magorrian} et~al.}{1998}]{Magorrian1998}
{Magorrian} J.,  et~al., 1998, \mn@doi [\aj] {10.1086/300353}, \href {https://ui.adsabs.harvard.edu/abs/1998AJ....115.2285M} {115, 2285}

\bibitem[\protect\citeauthoryear{{Maiolino} et~al.,}{{Maiolino} et~al.}{2023a}]{Maiolino2023b}
{Maiolino} R.,  et~al., 2023a, \mn@doi [arXiv e-prints] {10.48550/arXiv.2305.12492}, \href {https://ui.adsabs.harvard.edu/abs/2023arXiv230512492M} {p. arXiv:2305.12492}

\bibitem[\protect\citeauthoryear{{Maiolino} et~al.,}{{Maiolino} et~al.}{2023b}]{Maiolino2023a}
{Maiolino} R.,  et~al., 2023b, \mn@doi [arXiv e-prints] {10.48550/arXiv.2308.01230}, \href {https://ui.adsabs.harvard.edu/abs/2023arXiv230801230M} {p. arXiv:2308.01230}

\bibitem[\protect\citeauthoryear{{Marconi} \& {Hunt}}{{Marconi} \& {Hunt}}{2003}]{Marconi2003}
{Marconi} A.,  {Hunt} L.~K.,  2003, \mn@doi [\apjl] {10.1086/375804}, \href {https://ui.adsabs.harvard.edu/abs/2003ApJ...589L..21M} {589, L21}

\bibitem[\protect\citeauthoryear{{Marinacci} et~al.,}{{Marinacci} et~al.}{2018}]{Marinacci2018}
{Marinacci} F.,  et~al., 2018, \mn@doi [\mnras] {10.1093/mnras/sty2206}, \href {https://ui.adsabs.harvard.edu/abs/2018MNRAS.480.5113M} {480, 5113}

\bibitem[\protect\citeauthoryear{{Marshall}, {Ni}, {Di Matteo}, {Wyithe}, {Wilkins}, {Croft}  \& {Kuusisto}}{{Marshall} et~al.}{2020}]{Marshall2020}
{Marshall} M.~A.,  {Ni} Y.,  {Di Matteo} T.,  {Wyithe} J. S.~B.,  {Wilkins} S.,  {Croft} R. A.~C.,   {Kuusisto} J.~K.,  2020, \mn@doi [\mnras] {10.1093/mnras/staa2982}, \href {https://ui.adsabs.harvard.edu/abs/2020MNRAS.499.3819M} {499, 3819}

\bibitem[\protect\citeauthoryear{{Matsuoka} et~al.,}{{Matsuoka} et~al.}{2016}]{Matsuoka2016}
{Matsuoka} Y.,  et~al., 2016, \mn@doi [\apj] {10.3847/0004-637X/828/1/26}, \href {https://ui.adsabs.harvard.edu/abs/2016ApJ...828...26M} {828, 26}

\bibitem[\protect\citeauthoryear{{Matthee} et~al.,}{{Matthee} et~al.}{2023}]{Matthee23}
{Matthee} J.,  et~al., 2023, \mn@doi [arXiv e-prints] {10.48550/arXiv.2306.05448}, \href {https://ui.adsabs.harvard.edu/abs/2023arXiv230605448M} {p. arXiv:2306.05448}

\bibitem[\protect\citeauthoryear{{McAlpine} et~al.,}{{McAlpine} et~al.}{2016}]{McAlpine2016}
{McAlpine} S.,  et~al., 2016, \mn@doi [Astronomy and Computing] {10.1016/j.ascom.2016.02.004}, \href {https://ui.adsabs.harvard.edu/abs/2016A&C....15...72M} {15, 72}

\bibitem[\protect\citeauthoryear{{McAlpine}, {Bower}, {Harrison}, {Crain}, {Schaller}, {Schaye}  \& {Theuns}}{{McAlpine} et~al.}{2017}]{McAlpine2017}
{McAlpine} S.,  {Bower} R.~G.,  {Harrison} C.~M.,  {Crain} R.~A.,  {Schaller} M.,  {Schaye} J.,   {Theuns} T.,  2017, \mn@doi [\mnras] {10.1093/mnras/stx658}, \href {https://ui.adsabs.harvard.edu/abs/2017MNRAS.468.3395M} {468, 3395}

\bibitem[\protect\citeauthoryear{{McAlpine}, {Harrison}, {Rosario}, {Alexander}, {Ellison}, {Johansson}  \& {Patton}}{{McAlpine} et~al.}{2020}]{McAlpine2020}
{McAlpine} S.,  {Harrison} C.~M.,  {Rosario} D.~J.,  {Alexander} D.~M.,  {Ellison} S.~L.,  {Johansson} P.~H.,   {Patton} D.~R.,  2020, \mn@doi [\mnras] {10.1093/mnras/staa1123}, \href {https://ui.adsabs.harvard.edu/abs/2020MNRAS.494.5713M} {494, 5713}

\bibitem[\protect\citeauthoryear{{McCarthy}, {Schaye}, {Bird}  \& {Le Brun}}{{McCarthy} et~al.}{2017}]{McCarthy2017}
{McCarthy} I.~G.,  {Schaye} J.,  {Bird} S.,   {Le Brun} A. M.~C.,  2017, \mn@doi [\mnras] {10.1093/mnras/stw2792}, \href {https://ui.adsabs.harvard.edu/abs/2017MNRAS.465.2936M} {465, 2936}

\bibitem[\protect\citeauthoryear{{McConnell} \& {Ma}}{{McConnell} \& {Ma}}{2013}]{McConnell2013}
{McConnell} N.~J.,  {Ma} C.-P.,  2013, \mn@doi [\apj] {10.1088/0004-637X/764/2/184}, \href {https://ui.adsabs.harvard.edu/abs/2013ApJ...764..184M} {764, 184}

\bibitem[\protect\citeauthoryear{{Meece}, {Voit}  \& {O'Shea}}{{Meece} et~al.}{2017}]{Meece2017}
{Meece} G.~R.,  {Voit} G.~M.,   {O'Shea} B.~W.,  2017, \mn@doi [\apj] {10.3847/1538-4357/aa6fb1}, \href {https://ui.adsabs.harvard.edu/abs/2017ApJ...841..133M} {841, 133}

\bibitem[\protect\citeauthoryear{{Merlin} et~al.,}{{Merlin} et~al.}{2019}]{Merlin2019}
{Merlin} E.,  et~al., 2019, \mn@doi [\mnras] {10.1093/mnras/stz2615}, \href {https://ui.adsabs.harvard.edu/abs/2019MNRAS.490.3309M} {490, 3309}

\bibitem[\protect\citeauthoryear{{Merloni}, {Heinz}  \& {di Matteo}}{{Merloni} et~al.}{2003}]{Merloni2003}
{Merloni} A.,  {Heinz} S.,   {di Matteo} T.,  2003, \mn@doi [\mnras] {10.1046/j.1365-2966.2003.07017.x}, \href {https://ui.adsabs.harvard.edu/abs/2003MNRAS.345.1057M} {345, 1057}

\bibitem[\protect\citeauthoryear{{Montgomery}, {Orchiston}  \& {Whittingham}}{{Montgomery} et~al.}{2009}]{Montgomery2009}
{Montgomery} C.,  {Orchiston} W.,   {Whittingham} I.,  2009, Journal of Astronomical History and Heritage, \href {https://ui.adsabs.harvard.edu/abs/2009JAHH...12...90M} {12, 90}

\bibitem[\protect\citeauthoryear{{Murray}, {Quataert}  \& {Thompson}}{{Murray} et~al.}{2005}]{Murray2005}
{Murray} N.,  {Quataert} E.,   {Thompson} T.~A.,  2005, \mn@doi [\apj] {10.1086/426067}, \href {https://ui.adsabs.harvard.edu/abs/2005ApJ...618..569M} {618, 569}

\bibitem[\protect\citeauthoryear{{Naab} \& {Ostriker}}{{Naab} \& {Ostriker}}{2017}]{Naab2017}
{Naab} T.,  {Ostriker} J.~P.,  2017, \mn@doi [\araa] {10.1146/annurev-astro-081913-040019}, \href {https://ui.adsabs.harvard.edu/abs/2017ARA&A..55...59N} {55, 59}

\bibitem[\protect\citeauthoryear{{Naiman} et~al.,}{{Naiman} et~al.}{2018}]{Naiman2018}
{Naiman} J.~P.,  et~al., 2018, \mn@doi [\mnras] {10.1093/mnras/sty618}, \href {https://ui.adsabs.harvard.edu/abs/2018MNRAS.477.1206N} {477, 1206}

\bibitem[\protect\citeauthoryear{{Narayan} \& {Yi}}{{Narayan} \& {Yi}}{1994}]{Narayan1994}
{Narayan} R.,  {Yi} I.,  1994, \mn@doi [\apjl] {10.1086/187381}, \href {https://ui.adsabs.harvard.edu/abs/1994ApJ...428L..13N} {428, L13}

\bibitem[\protect\citeauthoryear{{Nelson} et~al.,}{{Nelson} et~al.}{2018}]{Nelson2018}
{Nelson} D.,  et~al., 2018, \mn@doi [\mnras] {10.1093/mnras/stx3040}, \href {https://ui.adsabs.harvard.edu/abs/2018MNRAS.475..624N} {475, 624}

\bibitem[\protect\citeauthoryear{{Ni}, {Di Matteo}, {Gilli}, {Croft}, {Feng}  \& {Norman}}{{Ni} et~al.}{2020}]{Ni2020}
{Ni} Y.,  {Di Matteo} T.,  {Gilli} R.,  {Croft} R. A.~C.,  {Feng} Y.,   {Norman} C.,  2020, \mn@doi [\mnras] {10.1093/mnras/staa1313}, \href {https://ui.adsabs.harvard.edu/abs/2020MNRAS.495.2135N} {495, 2135}

\bibitem[\protect\citeauthoryear{{Ni} et~al.,}{{Ni} et~al.}{2022}]{Ni2022}
{Ni} Y.,  et~al., 2022, \mn@doi [\mnras] {10.1093/mnras/stac351}, \href {https://ui.adsabs.harvard.edu/abs/2022MNRAS.513..670N} {513, 670}

\bibitem[\protect\citeauthoryear{{Penrose}}{{Penrose}}{1965}]{Penrose1965}
{Penrose} R.,  1965, \mn@doi [\prl] {10.1103/PhysRevLett.14.57}, \href {https://ui.adsabs.harvard.edu/abs/1965PhRvL..14...57P} {14, 57}

\bibitem[\protect\citeauthoryear{{Penrose} \& {Floyd}}{{Penrose} \& {Floyd}}{1971}]{Penrose1971}
{Penrose} R.,  {Floyd} R.~M.,  1971, \mn@doi [Nature Physical Science] {10.1038/physci229177a0}, \href {https://ui.adsabs.harvard.edu/abs/1971NPhS..229..177P} {229, 177}

\bibitem[\protect\citeauthoryear{{Pillepich} et~al.,}{{Pillepich} et~al.}{2018a}]{Pillepich2018a}
{Pillepich} A.,  et~al., 2018a, \mn@doi [\mnras] {10.1093/mnras/stx2656}, \href {https://ui.adsabs.harvard.edu/abs/2018MNRAS.473.4077P} {473, 4077}

\bibitem[\protect\citeauthoryear{{Pillepich} et~al.,}{{Pillepich} et~al.}{2018b}]{Pillepich2018b}
{Pillepich} A.,  et~al., 2018b, \mn@doi [\mnras] {10.1093/mnras/stx3112}, \href {https://ui.adsabs.harvard.edu/abs/2018MNRAS.475..648P} {475, 648}

\bibitem[\protect\citeauthoryear{{Planck Collaboration} et~al.,}{{Planck Collaboration} et~al.}{2014}]{planck_collaboration_2014}
{Planck Collaboration} et~al., 2014, \mn@doi [A\&A] {10.1051/0004-6361/201321529}, 571, A1

\bibitem[\protect\citeauthoryear{{Portegies Zwart}, {Baumgardt}, {Hut}, {Makino}  \& {McMillan}}{{Portegies Zwart} et~al.}{2004}]{PortegiesZwart2004}
{Portegies Zwart} S.~F.,  {Baumgardt} H.,  {Hut} P.,  {Makino} J.,   {McMillan} S. L.~W.,  2004, \mn@doi [\nat] {10.1038/nature02448}, \href {https://ui.adsabs.harvard.edu/abs/2004Natur.428..724P} {428, 724}

\bibitem[\protect\citeauthoryear{{Puchwein}, {Haardt}, {Haehnelt}  \& {Madau}}{{Puchwein} et~al.}{2019}]{Puchwein2019}
{Puchwein} E.,  {Haardt} F.,  {Haehnelt} M.~G.,   {Madau} P.,  2019, \mn@doi [\mnras] {10.1093/mnras/stz222}, \href {https://ui.adsabs.harvard.edu/abs/2019MNRAS.485...47P} {485, 47}

\bibitem[\protect\citeauthoryear{{Qin} et~al.,}{{Qin} et~al.}{2017}]{Qin2017}
{Qin} Y.,  et~al., 2017, \mn@doi [\mnras] {10.1093/mnras/stx1909}, \href {https://ui.adsabs.harvard.edu/abs/2017MNRAS.472.2009Q} {472, 2009}

\bibitem[\protect\citeauthoryear{{Rees}}{{Rees}}{1984}]{Rees1984}
{Rees} M.~J.,  1984, \mn@doi [\araa] {10.1146/annurev.aa.22.090184.002351}, \href {https://ui.adsabs.harvard.edu/abs/1984ARA&A..22..471R} {22, 471}

\bibitem[\protect\citeauthoryear{{Regan} \& {Haehnelt}}{{Regan} \& {Haehnelt}}{2009}]{Regan2009}
{Regan} J.~A.,  {Haehnelt} M.~G.,  2009, \mn@doi [\mnras] {10.1111/j.1365-2966.2009.14579.x}, \href {https://ui.adsabs.harvard.edu/abs/2009MNRAS.396..343R} {396, 343}

\bibitem[\protect\citeauthoryear{{Richstone} et~al.,}{{Richstone} et~al.}{1998}]{Richstone1998}
{Richstone} D.,  et~al., 1998, \mn@doi [\nat] {10.48550/arXiv.astro-ph/9810378}, \href {https://ui.adsabs.harvard.edu/abs/1998Natur.395A..14R} {385, A14}

\bibitem[\protect\citeauthoryear{{Ricotti} \& {Ostriker}}{{Ricotti} \& {Ostriker}}{2004}]{Ricotti2004}
{Ricotti} M.,  {Ostriker} J.~P.,  2004, \mn@doi [\mnras] {10.1111/j.1365-2966.2004.07942.x}, \href {https://ui.adsabs.harvard.edu/abs/2004MNRAS.352..547R} {352, 547}

\bibitem[\protect\citeauthoryear{{Robertson}}{{Robertson}}{2022}]{Robertson2022}
{Robertson} B.~E.,  2022, \mn@doi [\araa] {10.1146/annurev-astro-120221-044656}, \href {https://ui.adsabs.harvard.edu/abs/2022ARA&A..60..121R} {60, 121}

\bibitem[\protect\citeauthoryear{{Rosas-Guevara} et~al.,}{{Rosas-Guevara} et~al.}{2015}]{Rosas-Guevara2015}
{Rosas-Guevara} Y.~M.,  et~al., 2015, \mn@doi [\mnras] {10.1093/mnras/stv2056}, \href {https://ui.adsabs.harvard.edu/abs/2015MNRAS.454.1038R} {454, 1038}

\bibitem[\protect\citeauthoryear{{Salpeter}}{{Salpeter}}{1964}]{Salpeter1964}
{Salpeter} E.~E.,  1964, \mn@doi [\apj] {10.1086/147973}, \href {https://ui.adsabs.harvard.edu/abs/1964ApJ...140..796S} {140, 796}

\bibitem[\protect\citeauthoryear{{Schaffer}}{{Schaffer}}{1979}]{Schaffer79}
{Schaffer} S.,  1979, \mn@doi [Journal for the History of Astronomy] {10.1177/002182867901000104}, \href {https://ui.adsabs.harvard.edu/abs/1979JHA....10...42S} {10, 42}

\bibitem[\protect\citeauthoryear{{Schaye}}{{Schaye}}{2004}]{Schaye2004}
{Schaye} J.,  2004, \mn@doi [\apj] {10.1086/421232}, \href {https://ui.adsabs.harvard.edu/abs/2004ApJ...609..667S} {609, 667}

\bibitem[\protect\citeauthoryear{Schaye et~al.,}{Schaye et~al.}{2015}]{schaye_eagle_2015}
Schaye J.,  et~al., 2015, \mn@doi [\mnras] {10.1093/mnras/stu2058}, 446, 521

\bibitem[\protect\citeauthoryear{{Schaye} et~al.,}{{Schaye} et~al.}{2023}]{Schaye2023}
{Schaye} J.,  et~al., 2023, \mn@doi [\mnras] {10.1093/mnras/stad2419}, \href {https://ui.adsabs.harvard.edu/abs/2023MNRAS.526.4978S} {526, 4978}

\bibitem[\protect\citeauthoryear{{Schmidt}}{{Schmidt}}{1963}]{Schmidt1963}
{Schmidt} M.,  1963, \mn@doi [\nat] {10.1038/1971040a0}, \href {https://ui.adsabs.harvard.edu/abs/1963Natur.197.1040S} {197, 1040}

\bibitem[\protect\citeauthoryear{{Shakura} \& {Sunyaev}}{{Shakura} \& {Sunyaev}}{1973}]{Shakura1973}
{Shakura} N.~I.,  {Sunyaev} R.~A.,  1973, \aap, \href {https://ui.adsabs.harvard.edu/abs/1973A&A....24..337S} {24, 337}

\bibitem[\protect\citeauthoryear{{Shen}, {Hopkins}, {Faucher-Gigu{\`e}re}, {Alexander}, {Richards}, {Ross}  \& {Hickox}}{{Shen} et~al.}{2020}]{Shen2020}
{Shen} X.,  {Hopkins} P.~F.,  {Faucher-Gigu{\`e}re} C.-A.,  {Alexander} D.~M.,  {Richards} G.~T.,  {Ross} N.~P.,   {Hickox} R.~C.,  2020, \mn@doi [\mnras] {10.1093/mnras/staa1381}, \href {https://ui.adsabs.harvard.edu/abs/2020MNRAS.495.3252S} {495, 3252}

\bibitem[\protect\citeauthoryear{{Sijacki}, {Springel}, {Di Matteo}  \& {Hernquist}}{{Sijacki} et~al.}{2007}]{Sijacki2007}
{Sijacki} D.,  {Springel} V.,  {Di Matteo} T.,   {Hernquist} L.,  2007, \mn@doi [\mnras] {10.1111/j.1365-2966.2007.12153.x}, \href {https://ui.adsabs.harvard.edu/abs/2007MNRAS.380..877S} {380, 877}

\bibitem[\protect\citeauthoryear{{Sijacki}, {Vogelsberger}, {Genel}, {Springel}, {Torrey}, {Snyder}, {Nelson}  \& {Hernquist}}{{Sijacki} et~al.}{2015}]{Sijacki2015}
{Sijacki} D.,  {Vogelsberger} M.,  {Genel} S.,  {Springel} V.,  {Torrey} P.,  {Snyder} G.~F.,  {Nelson} D.,   {Hernquist} L.,  2015, \mn@doi [\mnras] {10.1093/mnras/stv1340}, \href {https://ui.adsabs.harvard.edu/abs/2015MNRAS.452..575S} {452, 575}

\bibitem[\protect\citeauthoryear{{Silk} \& {Rees}}{{Silk} \& {Rees}}{1998}]{Silk1998}
{Silk} J.,  {Rees} M.~J.,  1998, \mn@doi [\aap] {10.48550/arXiv.astro-ph/9801013}, \href {https://ui.adsabs.harvard.edu/abs/1998A&A...331L...1S} {331, L1}

\bibitem[\protect\citeauthoryear{{Somerville} \& {Dav{\'e}}}{{Somerville} \& {Dav{\'e}}}{2015}]{Somerville2015}
{Somerville} R.~S.,  {Dav{\'e}} R.,  2015, \mn@doi [\araa] {10.1146/annurev-astro-082812-140951}, \href {https://ui.adsabs.harvard.edu/abs/2015ARA&A..53...51S} {53, 51}

\bibitem[\protect\citeauthoryear{{Springel}}{{Springel}}{2005}]{Springel2005b}
{Springel} V.,  2005, \mn@doi [\mnras] {10.1111/j.1365-2966.2005.09655.x}, \href {https://ui.adsabs.harvard.edu/abs/2005MNRAS.364.1105S} {364, 1105}

\bibitem[\protect\citeauthoryear{{Springel}, {Di Matteo}  \& {Hernquist}}{{Springel} et~al.}{2005}]{Springel2005}
{Springel} V.,  {Di Matteo} T.,   {Hernquist} L.,  2005, \mn@doi [\mnras] {10.1111/j.1365-2966.2005.09238.x}, \href {https://ui.adsabs.harvard.edu/abs/2005MNRAS.361..776S} {361, 776}

\bibitem[\protect\citeauthoryear{{Springel} et~al.,}{{Springel} et~al.}{2018}]{Springel2018}
{Springel} V.,  et~al., 2018, \mn@doi [\mnras] {10.1093/mnras/stx3304}, \href {https://ui.adsabs.harvard.edu/abs/2018MNRAS.475..676S} {475, 676}

\bibitem[\protect\citeauthoryear{{Springel}, {Pakmor}, {Zier}  \& {Reinecke}}{{Springel} et~al.}{2021}]{Springel2021}
{Springel} V.,  {Pakmor} R.,  {Zier} O.,   {Reinecke} M.,  2021, \mn@doi [\mnras] {10.1093/mnras/stab1855}, \href {https://ui.adsabs.harvard.edu/abs/2021MNRAS.506.2871S} {506, 2871}

\bibitem[\protect\citeauthoryear{{Stanway} \& {Eldridge}}{{Stanway} \& {Eldridge}}{2018}]{BPASS2.2.1}
{Stanway} E.~R.,  {Eldridge} J.~J.,  2018, \mn@doi [\mnras] {10.1093/mnras/sty1353}, \href {https://ui.adsabs.harvard.edu/abs/2018MNRAS.479...75S} {479, 75}

\bibitem[\protect\citeauthoryear{{Tenneti}, {Di Matteo}, {Croft}, {Garcia}  \& {Feng}}{{Tenneti} et~al.}{2018}]{Tenneti2018}
{Tenneti} A.,  {Di Matteo} T.,  {Croft} R.,  {Garcia} T.,   {Feng} Y.,  2018, \mn@doi [\mnras] {10.1093/mnras/stx2788}, \href {https://ui.adsabs.harvard.edu/abs/2018MNRAS.474..597T} {474, 597}

\bibitem[\protect\citeauthoryear{{Tremaine} et~al.,}{{Tremaine} et~al.}{2002}]{Tremaine2002}
{Tremaine} S.,  et~al., 2002, \mn@doi [\apj] {10.1086/341002}, \href {https://ui.adsabs.harvard.edu/abs/2002ApJ...574..740T} {574, 740}

\bibitem[\protect\citeauthoryear{{Tremmel}, {Governato}, {Volonteri}  \& {Quinn}}{{Tremmel} et~al.}{2015}]{Tremmel15}
{Tremmel} M.,  {Governato} F.,  {Volonteri} M.,   {Quinn} T.~R.,  2015, \mn@doi [\mnras] {10.1093/mnras/stv1060}, \href {https://ui.adsabs.harvard.edu/abs/2015MNRAS.451.1868T} {451, 1868}

\bibitem[\protect\citeauthoryear{{Tremmel}, {Karcher}, {Governato}, {Volonteri}, {Quinn}, {Pontzen}, {Anderson}  \& {Bellovary}}{{Tremmel} et~al.}{2017}]{Tremmel2017}
{Tremmel} M.,  {Karcher} M.,  {Governato} F.,  {Volonteri} M.,  {Quinn} T.~R.,  {Pontzen} A.,  {Anderson} L.,   {Bellovary} J.,  2017, \mn@doi [\mnras] {10.1093/mnras/stx1160}, \href {https://ui.adsabs.harvard.edu/abs/2017MNRAS.470.1121T} {470, 1121}

\bibitem[\protect\citeauthoryear{{{\"U}bler} et~al.,}{{{\"U}bler} et~al.}{2023}]{Ubler2023}
{{\"U}bler} H.,  et~al., 2023, \mn@doi [\aap] {10.1051/0004-6361/202346137}, \href {https://ui.adsabs.harvard.edu/abs/2023A&A...677A.145U} {677, A145}

\bibitem[\protect\citeauthoryear{{Vijayan}, {Lovell}, {Wilkins}, {Thomas}, {Barnes}, {Irodotou}, {Kuusisto}  \& {Roper}}{{Vijayan} et~al.}{2021}]{FLARES-II}
{Vijayan} A.~P.,  {Lovell} C.~C.,  {Wilkins} S.~M.,  {Thomas} P.~A.,  {Barnes} D.~J.,  {Irodotou} D.,  {Kuusisto} J.,   {Roper} W.~J.,  2021, \mn@doi [\mnras] {10.1093/mnras/staa3715}, \href {https://ui.adsabs.harvard.edu/abs/2021MNRAS.501.3289V} {501, 3289}

\bibitem[\protect\citeauthoryear{{Virtanen} et~al.,}{{Virtanen} et~al.}{2020}]{scipy}
{Virtanen} P.,  et~al., 2020, \mn@doi [Nature Methods] {https://doi.org/10.1038/s41592-019-0686-2}, \href {https://rdcu.be/b08Wh} {17, 261}

\bibitem[\protect\citeauthoryear{{Vogelsberger} et~al.,}{{Vogelsberger} et~al.}{2014a}]{Vogelsberger2014b}
{Vogelsberger} M.,  et~al., 2014a, \mn@doi [\mnras] {10.1093/mnras/stu1536}, \href {https://ui.adsabs.harvard.edu/abs/2014MNRAS.444.1518V} {444, 1518}

\bibitem[\protect\citeauthoryear{{Vogelsberger} et~al.,}{{Vogelsberger} et~al.}{2014b}]{Vogelsberger2014a}
{Vogelsberger} M.,  et~al., 2014b, \mn@doi [\nat] {10.1038/nature13316}, \href {https://ui.adsabs.harvard.edu/abs/2014Natur.509..177V} {509, 177}

\bibitem[\protect\citeauthoryear{{Vogelsberger}, {Marinacci}, {Torrey}  \& {Puchwein}}{{Vogelsberger} et~al.}{2020}]{Vogelsberger2020}
{Vogelsberger} M.,  {Marinacci} F.,  {Torrey} P.,   {Puchwein} E.,  2020, \mn@doi [Nature Reviews Physics] {10.1038/s42254-019-0127-2}, \href {https://ui.adsabs.harvard.edu/abs/2020NatRP...2...42V} {2, 42}

\bibitem[\protect\citeauthoryear{{Volonteri} \& {Rees}}{{Volonteri} \& {Rees}}{2005}]{Volonteri2005}
{Volonteri} M.,  {Rees} M.~J.,  2005, \mn@doi [\apj] {10.1086/466521}, \href {https://ui.adsabs.harvard.edu/abs/2005ApJ...633..624V} {633, 624}

\bibitem[\protect\citeauthoryear{{Volonteri}, {Dubois}, {Pichon}  \& {Devriendt}}{{Volonteri} et~al.}{2016}]{Volonteri2016}
{Volonteri} M.,  {Dubois} Y.,  {Pichon} C.,   {Devriendt} J.,  2016, \mn@doi [\mnras] {10.1093/mnras/stw1123}, \href {https://ui.adsabs.harvard.edu/abs/2016MNRAS.460.2979V} {460, 2979}

\bibitem[\protect\citeauthoryear{{Volonteri}, {Habouzit}  \& {Colpi}}{{Volonteri} et~al.}{2021}]{Volonteri2021}
{Volonteri} M.,  {Habouzit} M.,   {Colpi} M.,  2021, \mn@doi [Nature Reviews Physics] {10.1038/s42254-021-00364-9}, \href {https://ui.adsabs.harvard.edu/abs/2021NatRP...3..732V} {3, 732}

\bibitem[\protect\citeauthoryear{{Weinberger} et~al.,}{{Weinberger} et~al.}{2017}]{Weinberger2017}
{Weinberger} R.,  et~al., 2017, \mn@doi [\mnras] {10.1093/mnras/stw2944}, \href {https://ui.adsabs.harvard.edu/abs/2017MNRAS.465.3291W} {465, 3291}

\bibitem[\protect\citeauthoryear{{Weinberger} et~al.,}{{Weinberger} et~al.}{2018}]{Weinberger2018}
{Weinberger} R.,  et~al., 2018, \mn@doi [\mnras] {10.1093/mnras/sty1733}, \href {https://ui.adsabs.harvard.edu/abs/2018MNRAS.479.4056W} {479, 4056}

\bibitem[\protect\citeauthoryear{{Wiersma}, {Schaye}  \& {Smith}}{{Wiersma} et~al.}{2009}]{Wiersma2009}
{Wiersma} R. P.~C.,  {Schaye} J.,   {Smith} B.~D.,  2009, \mn@doi [\mnras] {10.1111/j.1365-2966.2008.14191.x}, \href {https://ui.adsabs.harvard.edu/abs/2009MNRAS.393...99W} {393, 99}

\bibitem[\protect\citeauthoryear{{Wilkins}, {Feng}, {Di Matteo}, {Croft}, {Lovell}  \& {Waters}}{{Wilkins} et~al.}{2017}]{Wilkins2017}
{Wilkins} S.~M.,  {Feng} Y.,  {Di Matteo} T.,  {Croft} R.,  {Lovell} C.~C.,   {Waters} D.,  2017, \mn@doi [\mnras] {10.1093/mnras/stx841}, \href {https://ui.adsabs.harvard.edu/abs/2017MNRAS.469.2517W} {469, 2517}

\bibitem[\protect\citeauthoryear{Zinger et~al.,}{Zinger et~al.}{2020}]{ILLUSTRIS-Zinger+2020}
Zinger E.,  et~al., 2020, \mn@doi [Monthly Notices of the Royal Astronomical Society] {10.1093/mnras/staa2607}, 499, 768

\bibitem[\protect\citeauthoryear{{van der Velden}}{{van der Velden}}{2020}]{cmasher}
{van der Velden} E.,  2020, \mn@doi [The Journal of Open Source Software] {10.21105/joss.02004}, \href {https://ui.adsabs.harvard.edu/abs/2020JOSS....5.2004V} {5, 2004}

\makeatother
\end{thebibliography}




\end{document}